\RequirePackage{lineno}
\documentclass[twocolumn,trackchanges]{aastex62}
\usepackage[utf8]{inputenc}

\usepackage{acronym,booktabs,multirow,cellspace,tabularx,xspace,mathtools}
\usepackage{breakurl}

\graphicspath{{./}}

\newcolumntype{Y}{>{\centering\arraybackslash}X}

\newacro{ADI}[ADI]{accretion disk instability}
\newacro{BH}[BH]{black hole \acused{BHBH} \acused{NSBH}}
\newacro{CBC}[CBC]{compact binary coalescence}
\newacro{BNS}[BNS]{binary neutron star \acused{NS}}
\newacro{BBH}[BBH]{binary black hole \acused{BH}}
\newacro{CSG}[CSG]{circular sine--Gaussian}
\newacro{GBM}[GBM]{Gamma-ray Burst Monitor}
\newacro{BAT}[BAT]{Burst Alert Telescope}
\newacro{GCN}[GCN]{Gamma-ray Coordinates Network}
\newacro{GRB}[GRB]{Gamma-ray burst}
\newacro{GW}[GW]{gravitational wave}
\newacro{NS}[NS]{neutron star \acused{NNS}\acused{BNS}}
\newacro{NSBH}[NSBH]{neutron star--black hole \acused{NS} \acused{BH}}
\newacro{O1}[O1]{the first observing run of Advanced LIGO and Advanced Virgo}
\newacro{O2}[O2]{the second observing run of Advanced LIGO and Advanced Virgo}
\newacro{O3}[O3]{the third observing run of Advanced LIGO and Advanced Virgo}
\newacro{O3a}[O3a]{the first part of the third observing run of Advanced LIGO and Advanced Virgo}
\newacro{O3b}[O3b]{the second part of the third observing run of Advanced LIGO and Advanced Virgo}
\newacro{O4}[O4]{the fourth observing run of Advanced LIGO and Advanced Virgo}
\newacro{SNR}[SNR]{signal-to-noise ratio}
\newacro{VALID}[VALID]{Vetting Automation and Literature Informed Database}

\newcommand{\PYGRB}{\texttt{PyGRB}\xspace}
\newcommand{\PYCBC}{\texttt{PyCBC}\xspace}
\newcommand{\Xpipeline}{\texttt{X-Pipeline}\xspace}
\newcommand{\Msun}{\ensuremath{M_{\sun}}\xspace}

\newcommand{\OThreeAStart}{{1 April 2019 15:00 UTC}\xspace}
\newcommand{\OThreeAEnd}{{1 October 2019 15:00 UTC}\xspace}
\newcommand{\OThreeBStart}{{1 November 2019 15:00 UTC}\xspace}
\newcommand{\OThreeBEnd}{{27 March 2020 17:00 UTC}\xspace}
\newcommand{\CBCsInOThreeB}{{35}\xspace}

\newcommand{\DoubleDF}{{\ensuremath{85.3\,\%}\xspace}}
\newcommand{\SingleDF}{{\ensuremath{96.6\,\%}\xspace}}

\newcommand{\nAllGRB}{{108}\xspace}
\newcommand{\nLongRun}{{89}\xspace}
\newcommand{\nAmbiguousRun}{{12}\xspace}
\newcommand{\nShortRun}{{7}\xspace}

\newcommand{\nBurst}{{86}\xspace}
\newcommand{\nRedshift}{{2}\xspace}

\newcommand{\pvalBurstLowest}{\ensuremath{7.95\times 10^{-3}}\xspace}
\newcommand{\nameBurstLowest}{GRB~200224B\xspace}
\newcommand{\pvalBurstOTwo}{{\ensuremath{0.75}}\xspace}
\newcommand{\pvalBurstOThreeA}{{\ensuremath{0.30}}\xspace}
\newcommand{\pvalBurstOOne}{{\ensuremath{0.75}}\xspace}

\newcommand{\pvalBurst}{{\ensuremath{0.76}}\xspace}
\newcommand{\DCSGSeventy}{{\ensuremath{166}}\xspace}
\newcommand{\DCSGHundred}{{\ensuremath{126}}\xspace}
\newcommand{\DCSGOneFifty}{{\ensuremath{92}}\xspace}
\newcommand{\DCSGThreeHundred}{{\ensuremath{42}}\xspace}
\newcommand{\DADIA}{{\ensuremath{34}}\xspace}
\newcommand{\DADIB}{{\ensuremath{140}}\xspace}
\newcommand{\DADIC}{{\ensuremath{54}}\xspace}
\newcommand{\DADID}{{\ensuremath{22}}\xspace}
\newcommand{\DADIE}{{\ensuremath{52}}\xspace}

\newcommand{\nCBC}{\ensuremath{17}\xspace}
\newcommand{\binompvalCBC}{\ensuremath{0.07}\xspace}
\newcommand{\binompvalCBCCut}{\ensuremath{0.68}\xspace}
\newcommand{\DBNS}{\ensuremath{149}\xspace}
\newcommand{\DNSBHGen}{\ensuremath{207}\xspace}
\newcommand{\DNSBHAli}{\ensuremath{257}\xspace}
\newcommand{\pvalCBCLowestFAP}{\ensuremath{1.08\times 10^{-2}}\xspace}
\newcommand{\nameCBCLowestFAP}{GRB~200129A\xspace}

\newcommand{\hms}[3]{#1\ensuremath{^{\mathrm{h}}}#2\ensuremath{^{\mathrm{m}}}#3\ensuremath{^{\mathrm{s}}}}
\newcommand{\dms}[2]{#1\ensuremath{^{\circ}}#2'}

\newcommand{\pvalCBCOTwo}{{\ensuremath{0.30}}\xspace}
\newcommand{\pvalCBCOThreeA}{{\ensuremath{0.43}}\xspace}
\newcommand{\pvalCBCOOne}{{\ensuremath{0.57}}\xspace}

\newcommand{\Lstar}{{\ensuremath{L_* = 2 \times 10^{52}~$erg$~$s$^{-1}}}\xspace}
\newcommand{\Lstarstar}{{\ensuremath{L_{**} = 5 \times 10^{49}~$erg$~$s$^{-1}}}\xspace}
\newcommand{\alphaL}{{\ensuremath{\alpha_L = 0.94}}\xspace}
\newcommand{\betaL}{{\ensuremath{\beta_L = 2}}\xspace}
\newcommand{\LBNSOTwo}{{\ensuremath{L_{\mathrm{GRB~170817A}}=(1.6 \pm 0.6) \times 10^{47}~\mathrm{erg}~\mathrm{s}^{-1}}}\xspace}
\newcommand{\nbGRBOOne}{{\ensuremath{20}}\xspace}
\newcommand{\nbGRBOTwo}{{\ensuremath{41}}\xspace}
\newcommand{\nbGRBOTwoTot}{{\ensuremath{42}}\xspace}
\newcommand{\nbGRBOThreeA}{{\ensuremath{32}}\xspace}
\newcommand{\nbGRBOThreeB}{{\ensuremath{17}}\xspace}
\newcommand{\gammaL}{{\ensuremath{\gamma_L = 0.28 \pm 0.45}}\xspace}
\newcommand{\JDROFour}{{\ensuremath{R^{\mathrm{O4}}_{\mathrm{GW-GRB}} = 1.04^{+0.26}_{-0.27}~\mathrm{yr}^{-1}}}\xspace}

\begin{document}

\title{Search for Gravitational Waves Associated with Gamma-Ray Bursts
  Detected by {\it Fermi} and {\it Swift} During the LIGO-Virgo Run
  O3b}

\author{R.~Abbott}
\affiliation{LIGO Laboratory, California Institute of Technology, Pasadena, CA 91125, USA}
\author{T.~D.~Abbott}
\affiliation{Louisiana State University, Baton Rouge, LA 70803, USA}
\author{F.~Acernese}
\affiliation{Dipartimento di Farmacia, Universit\`a di Salerno, I-84084 Fisciano, Salerno, Italy}
\affiliation{INFN, Sezione di Napoli, Complesso Universitario di Monte S. Angelo, I-80126 Napoli, Italy}
\author{K.~Ackley}
\affiliation{OzGrav, School of Physics \& Astronomy, Monash University, Clayton 3800, Victoria, Australia}
\author{C.~Adams}
\affiliation{LIGO Livingston Observatory, Livingston, LA 70754, USA}
\author{N.~Adhikari}
\affiliation{University of Wisconsin-Milwaukee, Milwaukee, WI 53201, USA}
\author{R.~X.~Adhikari}
\affiliation{LIGO Laboratory, California Institute of Technology, Pasadena, CA 91125, USA}
\author{V.~B.~Adya}
\affiliation{OzGrav, Australian National University, Canberra, Australian Capital Territory 0200, Australia}
\author{C.~Affeldt}
\affiliation{Max Planck Institute for Gravitational Physics (Albert Einstein Institute), D-30167 Hannover, Germany}
\affiliation{Leibniz Universit\"at Hannover, D-30167 Hannover, Germany}
\author{D.~Agarwal}
\affiliation{Inter-University Centre for Astronomy and Astrophysics, Pune 411007, India}
% ...

\collaboration{The LIGO Scientific Collaboration, the Virgo Collaboration, and the KAGRA Collaboration}

\correspondingauthor{LIGO Scientific Collaboration, Virgo Collaboration, \& KAGRA Collaboration Spokespersons}
\email{lsc-spokesperson@ligo.org, virgo-spokesperson@ego-gw.it}

\date{\today}
\begin{abstract}

  We search for gravitational-wave signals associated with
  gamma-ray bursts detected by the {\it Fermi} and {\it Swift}
  satellites during the second half of the third observing run of
  Advanced LIGO and Advanced Virgo (\OThreeBStart--\OThreeBEnd).
  We conduct two independent searches: a generic gravitational-wave transients search to analyze
   \nBurst gamma-ray bursts and an analysis to target
   binary mergers with at least one neutron star as short gamma-ray burst progenitors for
  \nCBC events.
  We find no significant evidence for gravitational-wave signals associated with any of these gamma-ray bursts. A weighted binomial test of the combined results finds no evidence for sub-threshold gravitational wave signals associated with this GRB ensemble either.
   We use several source types and signal morphologies during the searches,
  resulting in lower bounds on the estimated distance to each gamma-ray burst.
  Finally, we constrain the population of low luminosity short gamma-ray bursts
  using results from the first to the third observing runs of Advanced LIGO and
  Advanced Virgo. The resulting population is in accordance with the local binary
  neutron star merger rate.
\end{abstract}

\section{Introduction}
\label{Introduction}

\aclp{GRB}\acused{GRB} \citep[\acp{GRB};][]{2015PhR...561....1K} are intense and highly
 variable flashes of gamma-rays (the prompt emission), followed by a long-lasting, multi-wavelength emission (the afterglow emission), typically observed in X-rays, optical, radio, and sometimes in gamma-rays. They are believed to be powered by ultra-relativistic jets produced by rapid accretion onto a central compact object: 
 a \aclu{BH} \citep[\ac{BH};][]{1993ApJ...405..273W,1999ApJ...518..356P} or a magnetar \citep{1998A&A...333L..87D,2001ApJ...552L..35Z}.

\acp{GRB} are divided into two classes, depending on the duration and the spectral hardness of the prompt emission \citep{1993ApJ...413L.101K}: long, soft \acp{GRB} (duration $\gtrsim$ 2 s) and short, hard \acp{GRB} (duration $<$ 2 s). 

Long \acp{GRB} are thought to be associated with the core collapse of massive stars. This connection is observationally supported by the identification of supernova (SN) signatures in a number of sufficiently close long \acp{GRB} \citep{1998Natur.395..670G,2003Natur.423..847H,2003ApJ...591L..17S}. Core-collapsing massive stars are also expected to emit \acp{GW}
if there is some asymmetry in the stellar-envelope ejection phase (\citealp{2006RPPh...69..971K,2009CQGra..26f3001O,2016PhRvD..93d2002G}). State-of-the-art models predict that such \ac{GW} radiation can be detected by current generation \ac{GW} interferometers only within our Galaxy \citep{2020PhRvD.101h4002A}; however, according to more extreme phenomenological models, such as long-lived bar-mode instabilities and disk fragmentation instabilities, \ac{GW} radiation could be detected even for extra-galactic sources \citep{2002ApJ...565..430F,2004PhRvD..69d4007V,2007ApJ...658.1173P,2009ApJ...702.1171C,2016PhRvD..93d2002G,2020PhRvD.101h4002A}.

Short \acp{GRB} were long believed to be associated with \ac{CBC}
composed by two neutron stars (NSs), a \ac{BNS} system, or a NS and a BH, a \ac{NSBH} binary \citep{1989Natur.340..126E,1991AcA....41..257P,1992ApJ...395L..83N}. The definitive proof of this association \citep{2017ApJ...848L..13A,2017ApJ...848L..12A} came with the joint detection of the \ac{BNS} merger \ac{GW} signal GW170817 \citep{2017PhRvL.119p1101A,2019PhRvX...9a1001A} and the  GRB 170817A \citep{2017ApJ...848L..15S,2017ApJ...848L..14G}. The ground-breaking electromagnetic follow-up campaign performed after this joint detection allowed the identification of the associated kilonova emission and of the \ac{GRB} afterglow emission (see \citealp{2017ApJ...848L..12A} and references therein). 

GRB 170817A was 2 to 6 orders of magnitude less energetic than other  \acp{GRB} \citep{2017ApJ...848L..13A}; the low luminosity of this source, together with the evolution of the X-ray and radio light curve \citep{2018MNRAS.478L..18T,2018ApJ...856L..18M,2018A&A...613L...1D} suggested an off-axis \ac{GRB} with a relativistic structured jet or a cocoon emission from the relativistic jet shocking its surrounding non-relativistic material. Subsequent very long baseline interferometry observations allowed constraints on the source size and its displacement, indicating that GW170817 produced a structured relativistic jet \citep{2019Sci...363..968G,2018Natur.561..355M}.

In \citet{2020arXiv201014550T} we presented targeted \ac{GW} follow-up of \acp{GRB} reported during \aclu{O3a} (\ac{O3a}; \OThreeAStart--\OThreeAEnd) by \emph{Fermi}'s Gamma-Ray Burst Monitor (\emph{Fermi}/GBM; \citealp{2009ApJ...702..791M}) and \emph{Swift}'s Burst Alert Telescope (\emph{Swift}/BAT; \citealp{2004ApJ...611.1005G,2005SSRv..120..143B,2020ApJ...900...35T}). No significant evidence for \ac{GW} signals associated with the \acp{GRB} that have been followed up has been found, nor for a population of unidentified sub-threshold signals.  

In this paper we present targeted \ac{GW} follow-up of \acp{GRB} reported during \ac{O3b} 
by \emph{Fermi}/GBM and \emph{Swift}/BAT. 
\ac{O3b} took place between \OThreeBStart and \OThreeBEnd. During \ac{O3b}, \CBCsInOThreeB CBC events have been identified with an inferred probability of astrophysical \ac{CBC} origin of $p_\mathrm{astro} > 0.5$  \citep{Abbott:2021catalogO3b}. 
The majority of them are classified as mergers of \ac{BBH} systems; however, several events are consistent with binary systems with at least one NS \citep{Abbott:2021catalogO3b}. One other event with lower $p_\mathrm{astro}$ was also published as a possible \ac{NSBH} coalescence \citep{nsbhO3}.
No EM counterparts have been reported so far in association with these events; however, given their large distances ($\gtrsim$ 300 Mpc) and their large error in the sky localization \citep{Abbott:2021catalogO3b}, it would have been difficult to detect an EM signal in association with these GW events.

In Section \ref{sec:grb_sample} we discuss the sample of \acp{GRB} analyzed in this paper. In Section \ref{sec:search_methods} we summarize the methods used to follow-up \acp{GRB}. In Section \ref{sec:results} we describe the results, and in Section \ref{sec:population} we present a population model analysis. Finally, in Section \ref{sec:conclusions} we present our concluding remarks.
\section{GRBs during O3b}
\label{sec:grb_sample}

Our \ac{GRB} sample consists of \nAllGRB events that occurred between \OThreeBStart 
and \OThreeBEnd. The vast majority of these events were identified in low-latency via 
notices circulated by the \ac{GCN} and subsequently refined with additional data from 
the \emph{Swift}/BAT catalog
and the \emph{Fermi}/GBM catalog.\footnote{\emph{Swift}/BAT Gamma-Ray Burst Catalog \href{https://swift.gsfc.nasa.gov/results/batgrbcat/}{swift.gsfc.nasa.gov/results/batgrbcat/}, and \emph{Fermi}/GBM Burst Catalog \href{https://heasarc.gsfc.nasa.gov/W3Browse/fermi/fermigbrst.html}{heasarc.gsfc.nasa.gov/W3Browse/fermi/fermigbrst.html}.} 
The \aclu{VALID}~\citep[\acs{VALID};][]{RCoyneThesis2015} is a dedicated 
processing system that tracks updates to the observed 
\ac{GRB} parameters, comparing time and localization data to ensure that the latest 
results are used for our GW analyses, and employing an automated literature search 
to identify particularly noteworthy events. 

We identify candidate events by classifying each \ac{GRB} as \emph{long}, 
\emph{short}, or \emph{ambiguous}. We classify events based on their $T_{90}$ 
(and its associated error $\delta T_{90}$), which is the time interval over 
which 90\% of the total background-subtracted photon counts are observed. 
\acp{GRB} are classified as \emph{short} when $T_{90} + |\delta T_{90}| < 2\,\mathrm{s}$, 
GRBs are classified as \emph{long} when $T_{90} - |\delta T_{90}| > 4\,\mathrm{s}$, 
and all remaining \acp{GRB} are labeled as \emph{ambiguous}. This long/short 
classification based on duration is only a general trend, and is not a perfect 
discriminator. For more robust classification one must also consider spectral 
properties, most commonly the spectral hardness or peak energy of the event, but 
since our sample consists of observations from multiple observatories with different 
spectral sensitivities we do not employ such quantities when organizing 
our sample. 

This classification process results in \nShortRun short GRBs, \nAmbiguousRun 
ambiguous \acp{GRB}, and \nLongRun long \acp{GRB}. Of all these GRBs, only \nRedshift have 
known redshifts:

\begin{itemize}
\item GRB 191221B \citep[$z=1.148$;][]{2019GCN.26553....1V, 2019GCN.26538....1K}
\item GRB 200205B \citep[$z=1.465$;][]{2020GCN.26998....1V}
\end{itemize}

In keeping with previous studies of this kind \citep{2017ApJ...841...89A, 2019ApJ...886...75A, 
2020arXiv201014550T}, we apply a generic transient search to all events, regardless of 
classification. In order to maximize our chances at identifying potential \ac{CBC} 
candidates, we apply our modeled search to all short and ambiguous \acp{GRB}. 
We also follow the same requirements on amount of data available within our network to process a given \ac{GRB}.
For the modeled search we select \acp{GRB} if there is a minimum amount of time in at least one detector around the time of the event. This gives us \nCBC events for our analysis corresponding with the observing time for the same selection criteria (\SingleDF~with at least one interferometer in observing mode).
For the generic transient search, we perform the selection by requiring enough data in at least two interferometers. This leads to \nBurst \acp{GRB} to analyze and is also compatible with the network observing time of at least 2 detectors (\DoubleDF).

\section{Search Methods} 
\label{sec:search_methods}

\subsection{Modeled search for compact binary mergers}
\label{sec:cbc-search}
This analysis is carried out by a coherent matched filtering pipeline, 
\PYGRB~\citep{Harry:2010fr, Williamson:2014wma}, contained within the open-source 
\PYCBC~\citep{alex_nitz_2020_3961510} suite which also relies heavily 
on the LALSuite~\citep{LALSuite} library. These searches seek to find 
candidate \ac{GW} signals coincident with the GRB triggers due to the inspiral 
and merger of \ac{BNS} or \ac{NSBH} binaries. We define a window around each 
GRB trigger, the \textit{on-source window}, which is $[-5, +1]$ s from the GRB 
trigger time. This window is based on the assumption that a GW may precede the 
prompt GRB emission by several seconds \citep{Lee_2007, Vedrenne_2009}, and 
was demonstrated by GW170817 \citep{Monitor:2017mdv}. The search also uses time 
surrounding the trigger, split into 6~s 
\textit{off-source windows}, to estimate the background. In total, the search 
uses $\sim$90 min of data around each GRB trigger to assign a significance to 
candidate events by ranking them against the background. 

The analysis requires a bank of template waveforms to carry out the matched filtering. 
We generated this bank using both geometric \citep{Brown_2012, Harry_2014} 
and stochastic methods~\citep{Harry_2008} for \ac{BNS} and \ac{NSBH} signals. 
The waveforms used in generating this bank are phenomenological 
inspiral--merger--ringdown waveform models of the IMRPhenomD family
\citep{Husa:2015iqa, Khan:2015jqa}. We choose to place limits on the bank, 
identical to those used in the \ac{O3a} template bank 
\citep{2020arXiv201014550T}, such that any \ac{NS} masses are limited to 
$[1.0,2.8]M_\odot$ and \ac{BH} masses are within $[2.8,25]M_\odot$. 
We conservatively set the mass cutoff between \ac{NS} and \ac{BH} based on an NS equation of state 
\citep{Kalogera_1996}. Functionally, this cutoff has no effect on the 
waveforms and is just used for nomenclature. The bank only contains aligned-spin \ac{BNS} and \ac{NSBH} 
binaries where the maximum dimensionless spin magnitude for \ac{NS}s is 
$0.05$ from the largest observed \ac{NS} spin in a binary \citep{Burgay_2003}. 
For \ac{BH}, we limit the spin to $0.998$ based on theory 
\citep{Thorne_1974}. Finally, we check to ensure that all potential binaries 
are viable GRB progenitors with the creation of an accretion disk 
able to power a GRB \citep{Pannarale:2014rea}. 

The only structural change between this bank and the bank used in the \ac{O3a} 
modeled searches \citep{2020arXiv201014550T} is the template placement for \ac{NSBH}
systems with total mass $M < 6 M_\odot$. Both banks are constructed by first performing a geometric generation
for a part of the parameter space. These templates are then seeded to a stochastic generation 
that fills the rest of the parameter space \citep{PhysRevD.93.124007}. The difference between 
the \ac{O3a} and \ac{O3b} banks is that the geometric generation for the \ac{O3a} bank extended 
through the low-mass \ac{NSBH} region whereas the \ac{O3b} bank limits the geometric generation 
to the BNS region. We made this change based on a bank verification which tests a bank's ability to recover a set of signals.
The result of this verification is a fitting-factor $(\mathcal{FF})$ that quantitatively 
measures the bank's performance \citep{PhysRevD.52.605}. The target for our template banks 
is to minimize the number of signals that have a $\mathcal{FF}$ less than a threshold, which we set at $0.97$ 
for our offline searches. For the same set of signals in the low-mass \ac{NSBH} region, 
the bank with a limited geometric generation recovers a factor of ten less signals with a 
fitting factor below $0.97$\textemdash when compared to the extended geometric bank. 
These results show that the limited geometric approach creates a more sensitive template bank for our searches.

\PYGRB uses this bank to rank candidate signals based on a re-weighted optimal 
SNR. This optimal SNR is the result of the coherent matched filter, and is 
re-weighted by how well the template matches the identified signal 
\citep{Harry:2010fr, Williamson:2014wma}. The search can then rank the significance 
of any event against the background using the \textit{off-source} windows. 
In order to improve this ranking statistic, we artificially increase the 
amount of off-source data by performing \textit{time slides} \citep{Williamson:2014wma}. 

To further determine the sensitivity of our searches, we inject signals  
into the off-source data and attempt to recover them. The signals that we 
choose to inject are generally in the same \ac{BNS} and \ac{NSBH} domains as 
the template bank, with a few important distinctions. Again, we replicate what 
was done in \ac{O3a} \citep{2020arXiv201014550T}, where the injected signals 
are split into three sets; a \ac{BNS} set with non-aligned (precessing) spins, 
an aligned-spin \ac{NSBH} set, and a precessing \ac{NSBH} set. The \ac{NS} masses 
in a \ac{BNS} binary are selected randomly from a normal distribution with a mean 
of $1.4M_\odot$ and variance of $0.2M_\odot$ \citep{Ozel_2012}. For \ac{NSBH} binaries, 
\ac{NS} masses are selected from a normal distribution with slightly more variance 
$(\mu=1.4M_\odot,\,\sigma=0.4M_\odot)$. The larger width reflects the greater 
uncertainty arising from a lack of observed NSBH systems. \ac{BH} masses are 
randomly selected from the following normal distribution 
$(\mu=10.0M_\odot,\,\sigma=6.0M_\odot)$. For all cases we place limits on the 
distributions similar to those used for the template bank. Randomly selected 
spin magnitudes are less than 0.4 for NSs based on the maximum observed pulsar spin 
\citep{Hessels:2006ze}, and less than 0.98 for BHs \citep{Miller:2014aaa}. 
For the two sets of injections that allow precessing signals, the orientations 
are also randomly selected. We also choose to use different waveform families 
than the ones used to generate the template bank to account for modeling 
uncertainty. We generate the \ac{BNS} injections using the SpinTaylorT2 
family, which are post-Newtonian approximations in the time domain
\citep{Sathyaprakash:1991mt, Blanchet:1996pi, Bohe:2013cla, Arun:2008kb, Mikoczi:2005dn, Bohe:2015ana, Mishra:2016whh}. 
The \ac{NSBH} sets both make use of the SEOBNRv3 family of waveforms. These 
waveforms are effective-one-body approximates that are tuned for precessing 
systems~\citep{Pan:2013rra, Taracchini:2013rva, Babak:2016tgq}. As with the 
template bank, we check to ensure that generated systems are capable GRB 
progenitors~\citep{Pannarale:2014rea}. These injection sets allow us to 
calculate the $90\%$ exclusion distance ($D_{90}$), which is 
the distance at which we recover $90\%$ of the injected signals 
with a significant ranking statistic.

\subsection{Search for generic GW transients}
\label{sec:Xpipeline-search}
This analysis, carried out with the \Xpipeline software  package \citep{2010NJPh...12e3034S, 2012PhRvD..86b2003W}, searches for excess power that is coherent across the GW detector network and consistent with the sky localization and time window of each GRB. Like the previous \Xpipeline analyses \citep{2017ApJ...841...89A, 2019ApJ...886...75A, 2020arXiv201014550T}, the
search time window starts 600\,s before the GRB trigger time and ends at 60\,s after trigger time, or $T_{90}$ after if $T_{90} > 60$~s. This is sufficient to cover the time delay between GW emission from a progenitor and any GRB prompt emission \citep{1995ApJ...452..145K, 2000ApJ...531L.119A, 2001ApJ...550..410M, 2003ApJ...586..356Z, 2005MNRAS.357..722L, 2007ApJ...670.1247W, 2008ApJ...685L..19B, 2009A&A...505..569B, 2009ApJ...700L..47L, 2009grb..book.....V}. While some GW emissions, such as from core-collapse SNe, are expected to reach frequencies up to a few kilohertz \citep{2019ApJ...876L...9R}, we restrict our search frequency range to the most sensitive band of the GW detectors, 20--500~Hz, since detecting such signals above a few hundred hertz requires extremely high GW energies~\citep[][Fig.~4]{2019PhRvD.100b4017A} and expanding the frequency range would also significantly increase the computational cost.

\Xpipeline produces time--frequency maps of the GW data coherently combined between the detectors. These maps give access to the temporal evolution of the spectral properties of the signal and enable the pipeline to search for clusters of pixels containing excess energy, referred to as \textit{events}. The pipeline assigns each event a detection statistic based on energy and ranks them accordingly. A coherent consistency test, based on correlations between data in different detectors, then vetoes events that are associated with noise transients. The surviving event with the largest ranking statistic is the best candidate for a GW detection, and the search quantifies its significance as the probability of the event being produced by the background alone. This is determined by comparing the SNR of the trigger within the 660\,s on-source window to the distribution of the SNRs of the loudest triggers in the 660\,s off-source windows. As a requirement, the off-source data consist of at least $\sim1.5$ hours of coincident data from at least two detectors around the time of a GRB. This is small enough to select data where the detectors should be in a similar state of operation as during the GRB on-source window, and large enough so that probability estimates using artificial time-shifting of the data are at the sub-percent level.

We quantify the sensitivity of the generic transient search by injecting simulated signals into off-source data. For each waveform family injected we determine the largest significance of any surviving cluster associated with the injections. We compute the percentage of injections that have a significance higher than the best event candidate and look for the amplitude at which this percentage is above 90\%, which sets the upper limit. We include O3b calibration errors~\citep{2021arXiv210700129S, 2021arXiv210703294V} by jittering the amplitude and arrival time according to a Gaussian distribution representative of the calibration uncertainties. As with the modeled search, these injection sets allow us to calculate 90\% exclusion distances.

We choose simulated waveforms to cover the search parameter space of three distinct sets of circular waveforms: BNS and NSBH binary inspiral signals, stellar collapse, and disk instability models. 

\begin{itemize}
\item Circular sine--Gaussian (CSG): signals representing GW emission from stellar collapses defined in Eq.~(1) of \citet{2017ApJ...841...89A} with a $Q$ factor of 9 and varying center frequency of 70~Hz, 100~Hz, 150~Hz, and 300~Hz. In all cases, we assume an optimistic emission of energy in GWs of $E_{\mathrm{GW}} = 10^{-2} M_{\odot} c^{2}$.
\item Binary inspiral: signals are characterized by a Gaussian distribution centered at $1.4 M_\odot$, with a width of $0.2 M_\odot$ for NS in a BNS, and with a width of $0.4 M_\odot$ for NS in NSBH. The distribution for GWs emitted by BNS mergers addresses the case of short GRB events as in \cite{2017ApJ...841...89A} and adopted in PyGRB search (Sec.~\ref{sec:cbc-search}).
\item Accretion disk instability (ADI): long-duration waveforms for GWs produced by instabilities in the magnetically suspended torus around a rapidly spinning BH. The model specifics and parameters used to generate the five families of ADI signals are the same as in the previous searches \citep{2017ApJ...841...89A, 2019ApJ...886...75A, 2020arXiv201014550T}.

\end{itemize}

In the O3a search, the sensitivity to long-duration ($\gtrsim10$\,s) signals was often limited by loud background noise transients known as \textit{glitches}~\citep{2021arXiv210111673D}. While \Xpipeline's coherent consistency tests easily veto these glitches, many long-duration simulated signals would overlap such a glitch by chance. In these cases the simulated signal and glitch would be clustered together and subsequently vetoed together. To address this problem, we implemented an \textit{autogating} procedure for O3b. For each detector, we compute the total energy in the whitened data stream over a 1\,s window. If this total fluctuates by more than 50 standard deviations above the median value, then the data is zeroed out over the interval where the threshold is exceeded and we apply an inverse 1\,s Tukey window at each end of the zeroed interval to transition smoothly between the whitened and zeroed data. To minimise the possibility of a loud GW transient triggering a gate, the procedure cancels a gate if there is a simultaneous energy excursion above 10 standard deviations in any other detector. The threshold of 50 standard deviations is low enough to gate the most problematic loud glitches, while being high enough that the only GWs zeroed out by the gate would have been detectable by all-sky searches. Empirically we find that this procedure is effective at reducing the impact of loud glitches without affecting the sensitivity to low-amplitude \ac{GW} signals.

For both search methods, we rank each candidate by calculating a $p$-value,
the probability of an event or a louder one in the on-source data,
given the background distribution, under the null hypothesis.
The $p$-value is calculated by counting the fraction of background trials
that contain an event with a greater signal-to-noise ratio than that of
the loudest on-source event.

\label{sec:burst-search}

\section{Results of analyses}
\label{sec:results}

\begin{figure}[!t]
  \begin{center}
    \includegraphics[width=\linewidth]{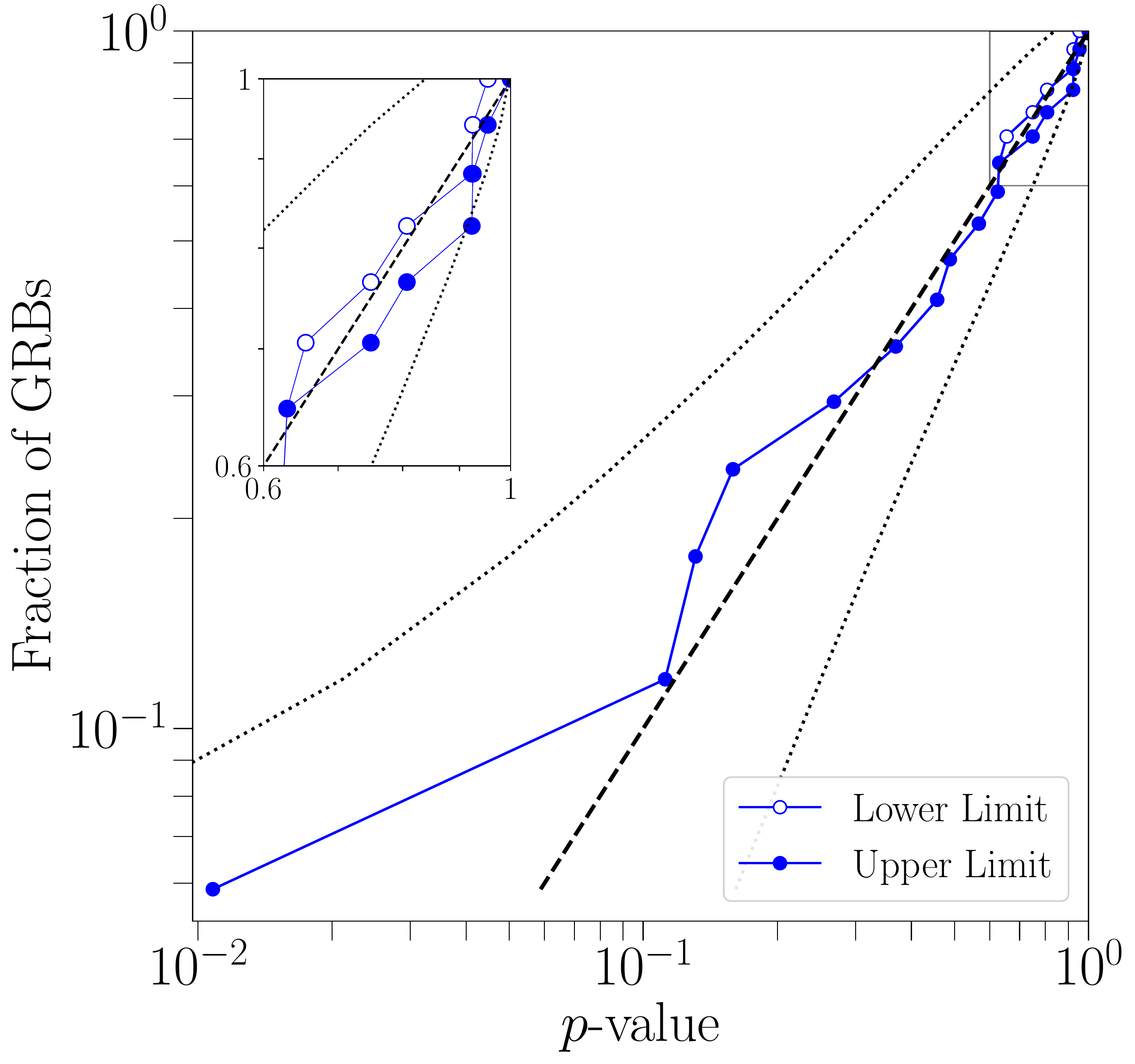}
  \end{center}
  \caption{The cumulative distribution of $p$-values for the loudest on-source 
    events for the modeled search in \ac{O3b}. If a trigger is found in the 
    on-source the upper and lower limits are identical to the reported $p$-value. 
    If no trigger is identified in the on-source window, we set an upper limit on 
    the $p$-value of 1, and a lower limit equal to the fraction of off-source 
    trials that also did not contain a trigger. The upper limits are plotted as 
    the curve with full circles and the lower limits are plotted as the 
    curve with empty circles. The dashed line indicates an expected uniform 
    distribution of $p$-values under a no-signal hypothesis, with the 
    corresponding $90\%$ band as the dotted lines.}
  \label{fig:pygrb-pvalue}
\end{figure}

\begin{figure}[!t]
  \centering
  \includegraphics[width=\linewidth]{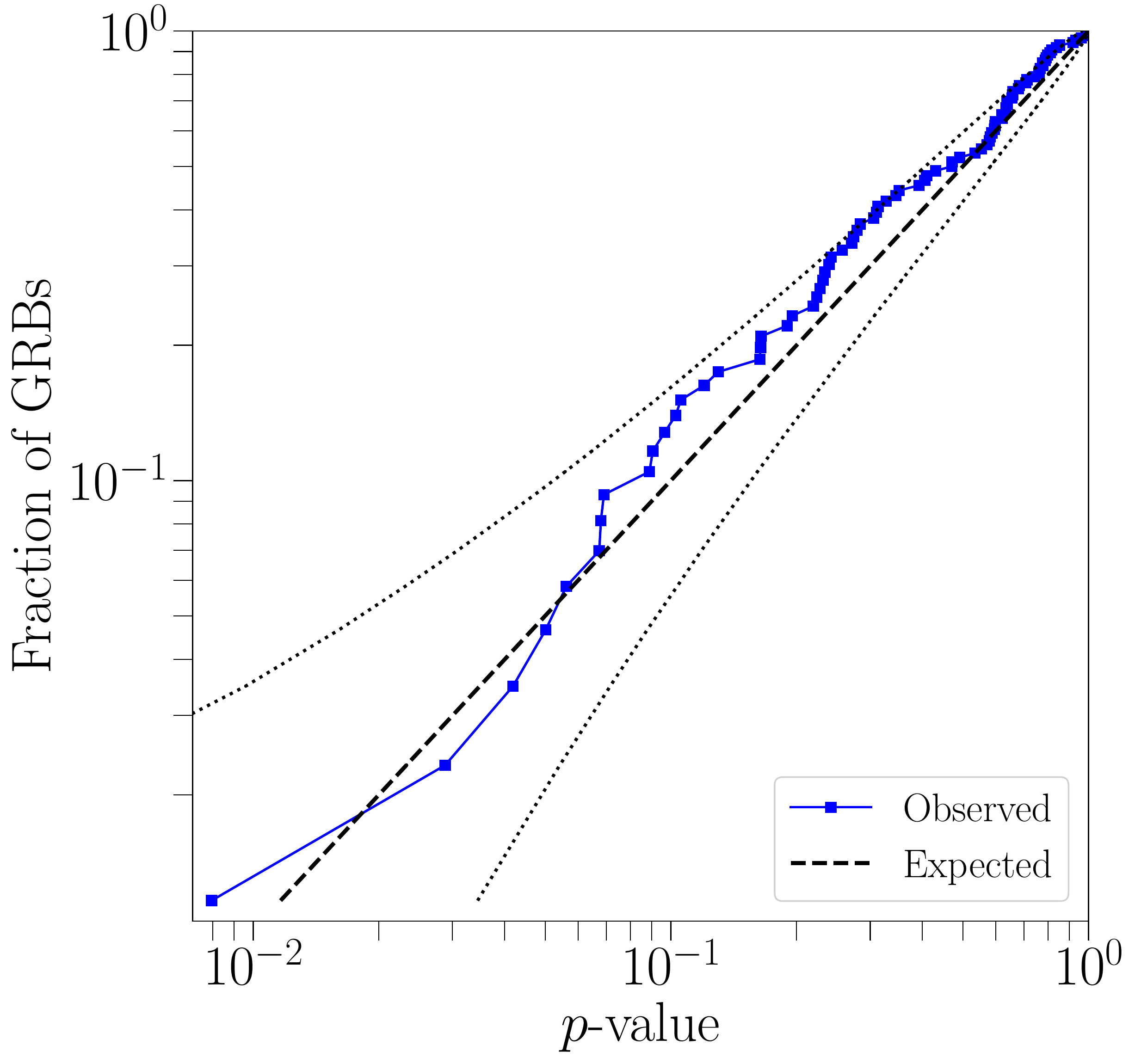}
  \caption{
	The cumulative distribution of $p$-values for the loudest on-source 
	    events for the generic transient search in \ac{O3b}.
	    Unlike with the modeled search, there is a $p$-value found
	    for all GRBs analyzed by the generic transient search,
	    so there are no upper and lower limits.
	    The observed values are plotted as the curve with full circles.
	    The dashed line indicates an expected uniform 
	    distribution of $p$-values under a no-signal hypothesis, with the 
	    corresponding $90\%$ band as the dotted lines.
		}
  \label{fig:burst-pvalue}
\end{figure}

We followed up \nBurst \ac{GRB} triggers with the generic 
transient method and \nCBC \acp{GRB} (those categorized as short or 
ambiguous) with the modeled search. None of the analyses indicate the 
presence of a statistically significant GW signal associated with one
or more of the GRBs. This null result is consistent with the 
estimated GW--GRB joint detection rate with \emph{Fermi}/GBM of 
$0.07$--$1.80$ per year reported previously in \citet{Authors:2019fue} for 
\ac{O2}.

We present the cumulative $p$-value distributions from both 
search methods in Figures \ref{fig:pygrb-pvalue} and \ref{fig:burst-pvalue}.
In these plots, a significant event would appear at a much lower $p$-value in the lower left
corner of the plots, and be outside (to the left) of the $90\%$ confidence region.
Both plots show that the $p$-value distributions are consistent with the background.

The most significant event from the modeled search had a $p$-value of 
\pvalCBCLowestFAP (\nameCBCLowestFAP). Through further investigation of this candidate event, a period of 
excess noise in one of the detectors was discovered $\sim~20$~s before the candidate time.
To determine the effect of this noise on the candidate, we used BayesWave 
to reconstruct the glitch and then clean the data by subtracting the reconstruction 
\citep{Cornish_2021, Pankow_2018}. After this cleaning, we conducted a coherent matched-filtering on the 
cleaned data and the recovered candidate was no longer significant with respect to
the background. This result suggests that much of the power of the candidate was 
caused by noise and not a \ac{GW}. Even if there is a quiet \ac{GW} at this time, it is not 
strong enough without the contribution from the glitch to survive ranking 
against the background in the analysis. 

The lowest reported $p$-value 
found during \ac{O3b} for the generic transient search was \pvalBurstLowest 
(\nameBurstLowest). Although this $p$-value is very small, it is not unexpected
given the high number of GRBs analyzed.

Given that no loud \ac{GW} signals were observed coincident with any of the \acp{GRB}
in either of our searches, we perform a weighted binomial test to determine the
probability of observing our set of $p$-values assuming a uniform background
distribution. A small probability would suggest that there may be a population 
of sub-threshold \ac{GW} signals that our searches did not identify. This type 
of weighted binomial test, fully described in the Appendix of \citet{2012ApJ...760...12A}, 
uses the lowest reweighted $p$-values from the searches. The resulting probability for the 
modeled search is \binompvalCBC. If we remove \nameCBCLowestFAP,
for which the small p-value is the result of noise, the probability 
becomes \binompvalCBCCut, suggesting no population of weak \ac{GW} signals. For the generic transient search,
the test gives a probability of \pvalBurst. These same weighted binomial tests
carried out in \ac{O3a} returned probabilities of \pvalCBCOThreeA and
\pvalBurstOThreeA for the modeled and generic transient searches, respectively
\citep{2020arXiv201014550T}. In \ac{O2} (removing GW170817/GRB 170817A) and \ac{O1} the probabilities were \pvalCBCOTwo and
\pvalBurstOTwo, and \pvalCBCOOne and \pvalBurstOOne, respectively
\citep{Authors:2019fue,2017ApJ...841...89A}. As in these previous analyses,
the probabilities obtained in \ac{O3b} suggest that no weak \acp{GW} can be 
attributed to the population of \acp{GRB}.

\begin{figure}[!t]
  \begin{center}
    \includegraphics[width=\linewidth]{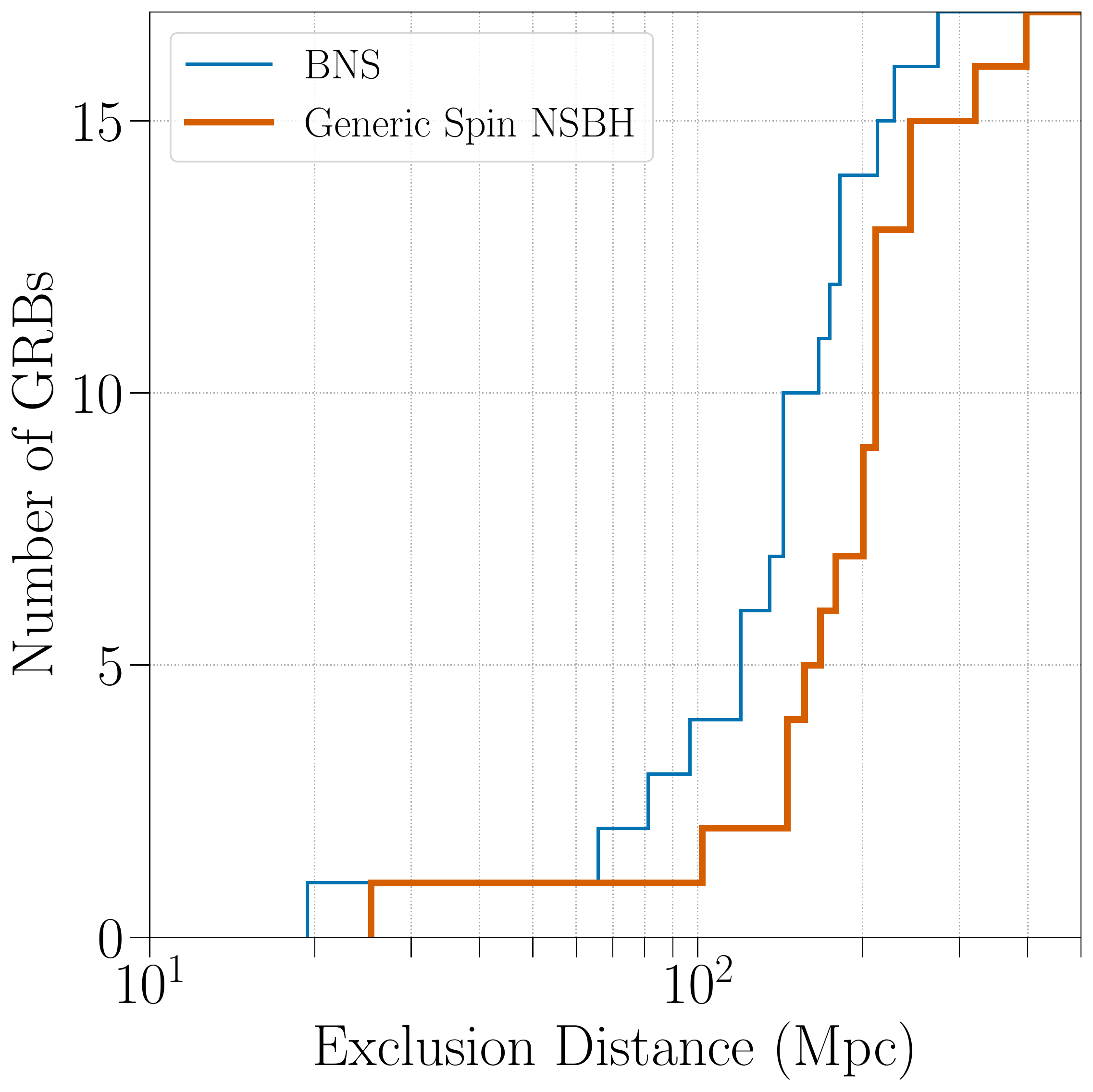}
  \end{center}
  \caption{Cumulative histograms of the 90\% exclusion
    distances, $D_{90}$, for the \nCBC GRBs that the modeled
    search followed up in \ac{O3b}. The thin blue line shows generically 
    spinning \ac{BNS} models and the thick orange line shows generically
    spinning \ac{NSBH} models.}
  \label{pygrb-90exclusion}
\end{figure}

\begin{figure}[!t]
  \centering
  \includegraphics[width=\linewidth]{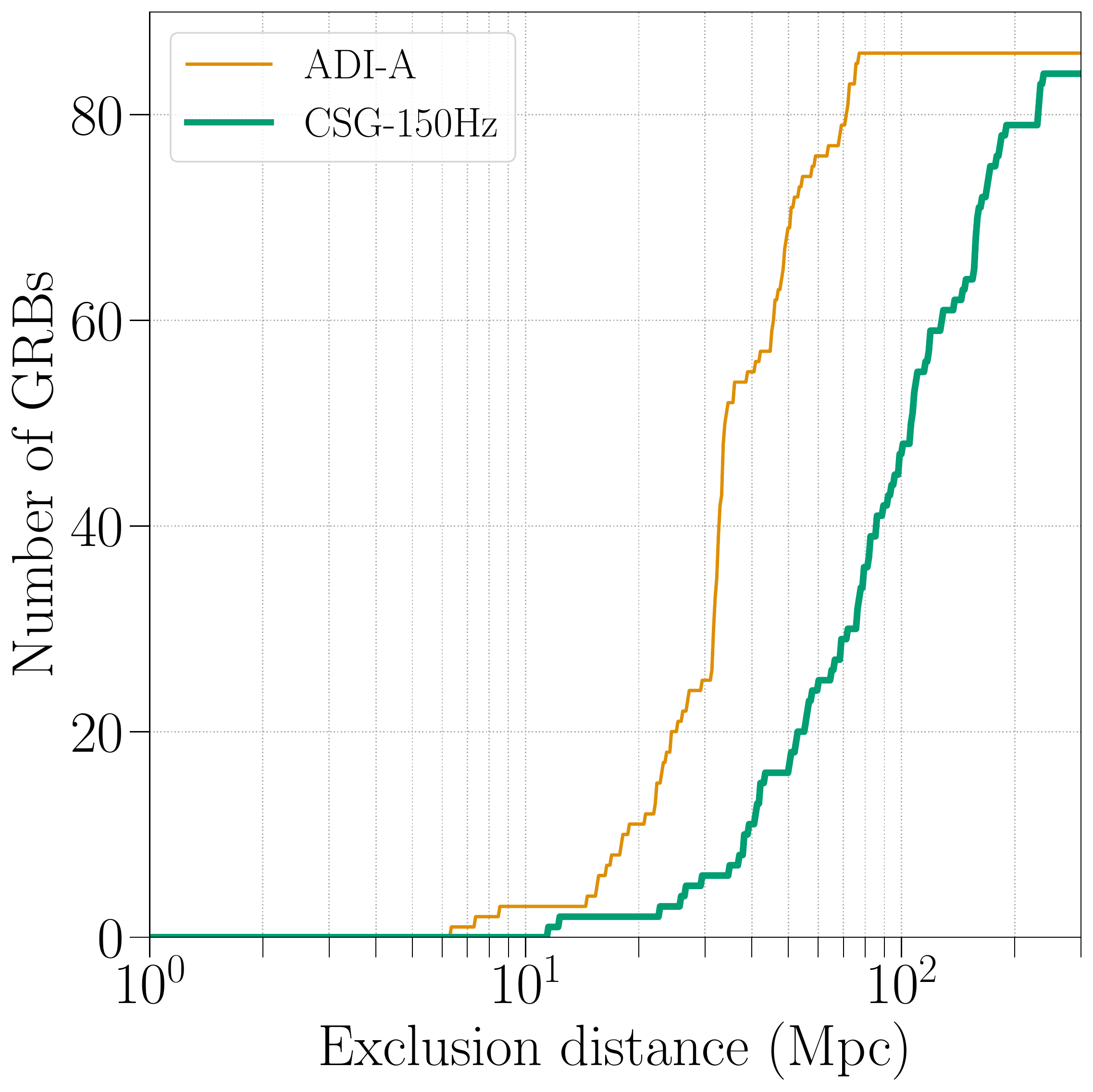}
  \caption{Cumulative histograms of the 90\%
    confidence exclusion distances, $D_{90}$, for accretion disk instability
    (\ac{ADI}) signal model A (orange, thin line) and circular sine--Gaussian
    (\ac{CSG}) 150 Hz model (green, thick line). For a given GRB and signal
    model, this is the distance within which 90\% of simulated signals inserted
    into off-source data are successfully recovered with a significance
    greater than the loudest on-source trigger.}
  \label{fig:burst-90exclusion}
\end{figure}

\begin{table}
  \hspace{0.5cm}
  \caption{\label{tab:exclDist} Median $90\%$ exclusion distances ($D_{90}$) for the both the 
    modeled and generic transient searches during \ac{O3b}. For the 
    modeled searches, we report the median ($D_{90}$) values for all three 
    simulated signal types. For the generic search, we report results obtained with 
    \acl{CSG}~\citep[\ac{CSG};][]{Abbott:2016cjt} and 
    \acl{ADI}~\citep[\ac{ADI};][]{vanPutten:2001gi,vanPutten:2014kja} 
    models.}
  \begin{tabularx}{\columnwidth}{ c c c c }
    \hline
    \hline
    \rule{0pt}{4ex}
    Modeled search &     & NSBH          & NSBH          \\
    (Short GRBs)   & BNS & Generic Spins & Aligned Spins \\
    \hline
    \rule[-2ex]{0pt}{4ex}
    $D_{90}$ [Mpc] & \DBNS & \DNSBHGen &  \DNSBHAli
  \end{tabularx}
  \begin{tabularx}{\columnwidth}{ c c c c c}
    \hline
    \hline
    \rule{0pt}{4ex}
    Generic transient search & CSG    & CSG     & CSG     & CSG     \\ 
    (All GRBs)       & 70\,Hz & 100\,Hz & 150\,Hz & 300\,Hz \\
    \hline
    \rule[-2ex]{0pt}{4ex}
    $D_{90}$ [Mpc] & \DCSGSeventy & \DCSGHundred & \DCSGOneFifty & \DCSGThreeHundred
  \end{tabularx}
  \begin{tabularx}{\columnwidth}{ c c c c c c }
    \hline
    \hline
    \rule{0pt}{4ex}
    Generic transient search & ADI & ADI & ADI & ADI & ADI \\
    (All GRBs)       & A   & B   & C   & D   & E   \\
    \hline
    \rule[-2ex]{0pt}{4ex}
    $D_{90}$ [Mpc] & \DADIA & \DADIB & \DADIC & \DADID & \DADIE \\
    \hline
  \end{tabularx}
\end{table}

In Fig.\,\ref{pygrb-90exclusion}, we present the cumulative $90\%$
exclusion distances for the \nCBC \acp{GRB} analyzed with the modeled 
search. The first of these \nCBC \acp{GRB}, GRB 200323A, has significantly 
lower exclusion distances than the rest. We can attribute this to the fact 
that the analysis of this GRB only used data from the Virgo interferometer. 
Furthermore, this GRB has a sub-optimal sky location for the Virgo 
interferometer with a sensitivity, when compared to an optimal 
sky-location, of $\sim 30\%$. Both of these factors produce the 
relatively small exclusion distances for the first step in the histogram. 
Table~\ref{tab:exclDist} reports the median $D_{90}$ for the \nCBC \acp{GRB} 
analyzed with the modeled search. It shows median values for all three
of the injected signal types described in Sec.\,\ref{sec:cbc-search}. For 
comparison, all three of these median values are $10$--$30\%$ larger than those
reported from the same modeled search in \ac{O3a} ~\citep{2020arXiv201014550T}. 
This difference stems from having a larger fraction of \acp{GRB} in \ac{O3b} that by chance 
arrived with better LIGO--Virgo antenna factors on average, bringing up the median values.
The individual $D_{90}$ values for each of the \nCBC \acp{GRB}
analyzed with the modeled search can be seen in Table~\ref{tab:combined}.

Similar to the modeled search, we derive 
a $90\%$ confidence level lower limit on the distance for each 
of the \nBurst \acp{GRB} analyzed with the generic transient search, based 
on the different emission models described in Sec.\,\ref{sec:burst-search}.
We present the distribution of $D_{90}$ values
for the \ac{ADI} model A~\citep{vanPutten:2001gi, vanPutten:2014kja}
and for a \ac{CSG} with central frequency of
$150$~Hz~\citep{Abbott:2016cjt} in Fig.~\ref{fig:burst-90exclusion}.
The limits reported depend on the sensitivity of the instruments
in the network, which change with time and sky localization 
of the GRB events. We marginalize these limits 
over errors introduced by detector calibration.
In Table~\ref{tab:exclDist}, we report the median exclusion distance
limits, $D_{90}$, for the set of GRBs for the different signals described
in Sec.\,\ref{sec:burst-search}. The limits vary by nearly an order
of magnitude due to the variety of signals used in our analysis. On average
the median values for the \ac{O3b} generic transient search are about $50\%$ greater
than those reported in \ac{O3a}~\citep{2020arXiv201014550T}.
We can primarily attribute this improvement to the use of autogating in
\ac{O3b}: the increase in exclusion distances is highest (up to a factor of two)
for the longest-duration waveforms, which are most impacted by the glitches
removed by autogating (as explained in Sec.\,\ref{sec:search_methods}).
The exclusion distances for the shorter-duration CSG waveforms, which are not
expected to be affected by autogating, increased by about 30\% on average.
This is more than could be accounted for by chance differences in
the LIGO--Virgo antenna factors between the two samples. Rather, the increase
is likely due to improvements in the performance of the detectors themselves,
such as through the reduction of noise caused by scattered light in the
LIGO detectors~\citep{Soni_2021} or the improvement in
sensitivity of the Virgo detector~\citep{2021arXiv210111673D}.
We report the $D_{90}$ values found for each
\ac{GRB} in the case of \ac{ADI} model A simulated signals and
\ac{CSG} simulated signals with central frequency of $150$~Hz in Table~\ref{tab:combined}, at the end of this paper.
\section{Population studies}
\label{sec:population}

We use the results obtained from the GW followup analysis of GRBs to put constraints on the low-luminosity short GRB population. For this purpose, we describe the short GRB population through a simple luminosity function model following~\citep{Wanderman_2015}, extended at low luminosities following the procedure described in~\citep{2019ApJ...886...75A}. We can then model the luminosity distribution through a power law with two breaks

\begin{equation}\label{eqn:lum_2breaks_pdf}
\begin{split}
\displaystyle \phi_0 (L_{\mathrm{iso}}) & \! \equiv \! \frac{\mathrm{d}P}{\mathrm{d} \log(L_{\mathrm{iso}})} \! = \! \\ &  \! = \! \begin{cases} 
\displaystyle \left( \frac{L_{\mathrm{iso}}}{L_{**}} \right)^{-\gamma_L} \! \left(\frac{L_{**}}{L_*} \right)^{-\alpha_L} \! , & \! \! \! L_0 \leq L_{\mathrm{iso}} \leq L_{**} \\
\displaystyle \left( \frac{L_{\mathrm{iso}}}{L_*} \right)^{-\alpha_L} \! , & \! \! \! L_{**} < L_{\mathrm{iso}} \leq L_*~, \\
\displaystyle \left( \frac{L_{\mathrm{iso}}}{L_*} \right)^{-\beta_L} \! , & \! \! \!  L_{\mathrm{iso}} > L_*
\end{cases}
\end{split}
\end{equation}
where $L_{\mathrm{iso}}$ is the isotropic equivalent GRB luminosity and for which we have \Lstar, \Lstarstar, \alphaL and \betaL~\citep{Wanderman_2015}. We do not take into account the measurement uncertainties for those fixed parameters as they would not significantly influence the analysis. The parameters on which we aim to put constraints using the joint GW--GRB analysis are the low-luminosity power index $\gamma_L$ and the low-luminosity cutoff for our population $L_0$. To make the dependence from these parameters clearer, we refer to the luminosity distribution as $\phi_0(L_{\mathrm{iso}}) \equiv \phi_0(L_{\mathrm{iso}}, \gamma_L, L_0)$. A Bayesian analysis constrains the parameters $\gamma_L$ and $L_0$ using the results from the O1, O2, O3a and O3b \PYGRB searches~\citep{Harry:2010fr, Williamson:2014wma, 2019ApJ...886...75A, 2020arXiv201014550T} and the results on BNS rates from~\citet{Abbott_2021}.

Under certain conditions, NSBH mergers can also produce sGRBs \citep{1992ApJ...395L..83N} and a small fraction of sGRBs can arise from local magnetar giant flares \citep{2021Burns_magnetar_giant_flares}. For simplicity, we ignore those relatively uncommon possibilities here.
We assume that BNS coalescences are the only progenitors for short GRBs, since there are restricted conditions under which an NSBH coalescence results into a short GRB \citep{Pannarale:2014rea}.

First, we compute the observed cumulative rate distribution $C_R^{\mathrm{obs}}(z, \gamma_L, L_0)$ as a function of redshift $z$, $\gamma_L$ and $L_0$. To do so, we take into account the cosmic rate density for short GRB explosions $\psi(z)$ adopting its form given in \citet{Wanderman_2015}. A Band function models the energy spectrum of the short GRBs \citep{1993ApJ...413..281B} with power indices $\alpha_{\mathrm{Band}} = -0.5$, $\beta_{\mathrm{Band}} = -2.25$ and peak energy $E_\mathrm{peak} = 800~\mathrm{keV}$, and we use Eq.~\eqref{eqn:lum_2breaks_pdf} as the luminosity distribution function for our population of short GRBs. As in \citet{Wanderman_2015}, we consider short GRBs detectable in gamma-rays when their 64 ms peak photon flux is above $P^{\mathrm{th}}_{64} = 2.37~\mathrm{photons}~\mathrm{cm}^{-2}~\mathrm{s}^{-1}$ in the energy window considered for \textit{Fermi}/GBM, i.e.\ $[50$--$300]~\mathrm{keV}$. We then compute the cumulative observed rate distribution as 
\begin{equation}\label{eqn:cum_obs_rate}
C_R^{\mathrm{obs}}(z, \gamma_L, L_0) = \int_{0}^{z} \frac{\mathrm{d}P^{\mathrm{GRB}}_{\mathrm{obs}}}{\mathrm{d}z'} \mathrm{d}z'~,
\end{equation}
where the differential probability of having an observed short GRB is defined as
\begin{equation}\label{eqn:diff_obs_rate}
\displaystyle \frac{\mathrm{d}P^{\mathrm{GRB}}_{\mathrm{obs}}}{\mathrm{d}z} \propto \frac{\psi(z)}{(1+z)} \frac{\mathrm{d}V}{\mathrm{d}z} \epsilon(z, \gamma_L, L_0)~.
\end{equation}
Here in Eq.~\eqref{eqn:diff_obs_rate}, $\psi(z)$ is the short GRB redshift distribution, $\mathrm{d}V/\mathrm{d}z$ is the differential comoving volume and $\epsilon(z, \gamma_L, L_0)$ is the efficiency curve for the \textit{Fermi}/GBM detector as a function of redshift and of the low-luminosity parameters of the luminosity distribution.

Using this rate distributon, we build a prior probability distribution function (PDF) $\Pi(\gamma_L, L_0)$. The prior is built starting from a flat distribution in the logarithms of the local observed rate density and of $L_0$, since those quantities can span over several orders of magnitude, then it is rescaled by the posterior cumulative distribution function of the BNS local rate density from~\citet{Abbott_2021}. This last factor formalizes the assumption that most of the short GRBs are produced in BNS coalescences.  For all computations, we consider a flat $\Lambda \mathrm{CDM}$ cosmology with $h_0 = 0.7$, $\Omega_{\mathrm{m}} = 0.3$ and $\Omega_{\Lambda} = 0.7$, in order to be consistent with the analysis done in \citet{Wanderman_2015}.

We define the likelihood function $\mathfrak{L}(x | \gamma_L, L_0)$ (where $x$ indicates our set of data) as the probability of detecting no GW transients associated with short or ambiguous GRBs during O1, O3a and O3b and of detecting one single GW transient associated to a GRB observed during the O2 run. Furthermore, we impose that the joint detection occured at the redshift measured for NGC~4993, the host galaxy of the event GW170817 \citep[$z_{\mathrm{NGC}~4993} = 0.009783$;][]{Levan_2017} and that the luminosity of the corresponding GRB is in the luminosity range measured for GRB~170817A \LBNSOTwo~\citep{2017ApJ...848L..13A}. For our purpose we use the set of GW efficiency curves computed through the \PYGRB analysis of the short and ambiguous GRBs events detected during the O1, O2, O3a and O3b runs (respectively \nbGRBOOne, \nbGRBOTwo, \nbGRBOThreeA and \nbGRBOThreeB events analyzed).\footnote{There are actually \nbGRBOTwoTot efficiency curves available from the O2 \PYGRB analysis, but the efficiency curve corresponding to GRB~170817A was not computed properly since the pipeline considered the GW170817 event as a background event.}

Given a detected GRB \textit{i} during O2, we compute the probability of a joint GW detection like the one observed during this run
\begin{equation}\label{eqn:prob_det_GRB_GW}
\begin{split}
\displaystyle P^{\mathrm{det}}_{i}(\gamma_L, L_0) & = \int_{0}^{\infty} \phi_0(L, \gamma_L, L_0) \ln \mathcal{N}_{\tilde{L}}(L) \mathrm{d}\ln{L} \\ & \times \int_{0}^{\infty} \eta_i(z) \frac{\mathrm{d}P^{\mathrm{GRB}}_{\mathrm{obs}}}{\mathrm{d}z} \delta(z-z_{\mathrm{NGC}~4993}) \mathrm{d}z~.
\end{split}
\end{equation}
Here $\eta_i(z)$ is the efficiency curve corresponding to the given GRB and ${\mathrm{d}P^{\mathrm{GRB}}_{\mathrm{obs}}}/{\mathrm{d}z}$ has been defined in Eq.~\eqref{eqn:diff_obs_rate}. In order to set the joint detection to have the same luminosity of GRB~170817A and the same redshift of GW170817, we choose $\mathcal{N}_{\tilde{L}}(L)$ to be a log-normal distribution with mean $L_{\mathrm{GRB}170817\mathrm{A}} = \tilde{L}$ with $\sigma_{\tilde{L}}$ being the error on the measurement of $\tilde{L}$, and we use a Dirac delta distribution $\delta(z-z_{\mathrm{NGC}~4993})$ because our analysis is insensitive to small variations in the assumed redshift.

Analogously, we can compute the probability of not having a joint GW detection associated to a given GRB detected during O1, O3a or O3b
\begin{equation}\label{eqn:prob_no_det_GRB_GW}
\displaystyle P^{\mathrm{no~det}}_{i}(\gamma_L, L_0) = 1 - \int_{0}^{\infty} \eta_i(z) \frac{\mathrm{d}P^{\mathrm{GRB}}_{\mathrm{obs}}}{\mathrm{d}z} \mathrm{d}z~.
\end{equation}

We then obtain that the probability of a single joint detection during O2 is
\begin{equation}\label{eqn: O2_jd_prob}
\begin{split}
\displaystyle \mathbf{P}_{\mathrm{O2}} & (\gamma_L, L_0) = \\ & = \sum_{i}^{N^{\mathrm{GRB}}_{\mathrm{O2}}} \Bigg( P^{\mathrm{det}}_{i}(\gamma_L, L_0) \prod_{j \neq i} P^{\mathrm{no~det}}_{j}(\gamma_L, L_0) \Bigg)
\end{split}
\end{equation}
while the probability of not having a joint detection during O1, O3a and O3b is
\begin{equation}\label{eqn: O3_jd_prob}
\begin{split}
\displaystyle \mathbf{P}_{\mathrm{O1+O3}} & (\gamma_L, L_0) = \! \\ & = \! \! \prod_{i}^{N^{\mathrm{GRB}}_{\mathrm{O1}}} \! \! \! P^{\mathrm{no~det}}_{i}(\gamma_L, L_0) \! \! \! \prod_{i}^{N^{\mathrm{GRB}}_{\mathrm{O3}}} \! \! \! P^{\mathrm{no~det}}_{i}(\gamma_L, L_0)
\end{split}
\end{equation}
then the obtained likelihood is 
\begin{equation}\label{eqn: basic_likelihood_complete}
\displaystyle \mathfrak{L}(x|\gamma_L, L_0) = \mathbf{P}_{\mathrm{O2}}(\gamma_L, L_0)\mathbf{P}_{\mathrm{O1+O3}}(\gamma_L, L_0)~.
\end{equation}

\begin{figure}[!t]
	\begin{center}
		\includegraphics[width=\linewidth]{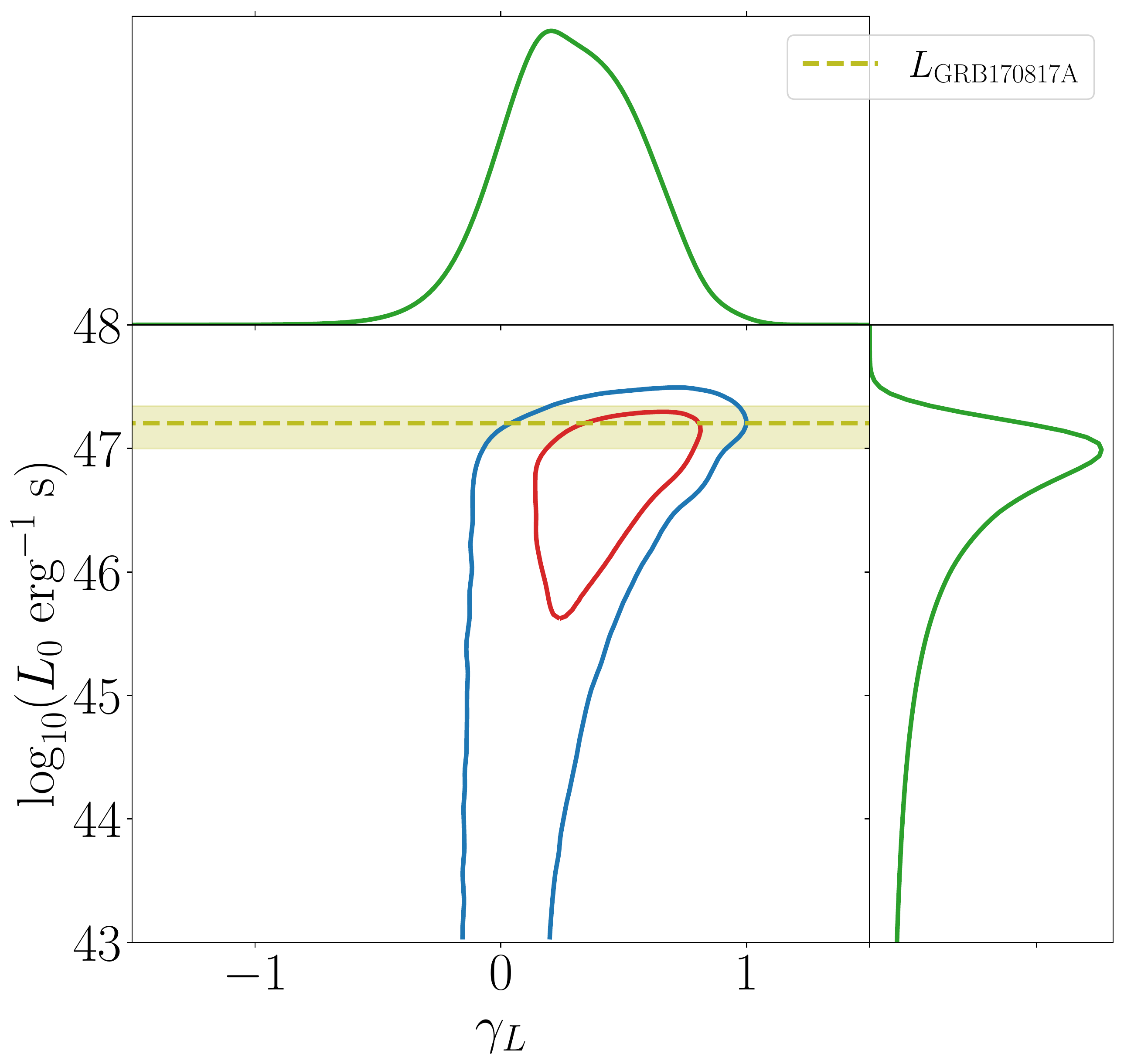}
	\end{center}
	\caption{Contour plot of the two-dimensional posterior as a function of the $\gamma_L$ parameter (x-axis) and of the base 10 logarithm of $L_0$ (y-axis) with plots of the corresponding marginalized posterior curves (in green). The contours correspond to the $90\%$ and $50\%$ credible regions (respectively in blue and red) for the two parameters. The bounds regions for those two parameters are compatible with the measured luminosity from GRB~170817A (yellow dashed line with shaded area) as its value is greater than $L_0$ for the bulk of the values of our population. The marginalized posterior for $L_0$ peaks around $L=L_{\mathrm{GRB~170817A}}$ because of the likelihood factor which requires that the joint detection happened around that value.}
	\label{fig:posteriorPDF}
\end{figure}

Finally, we compute the posterior $P(\gamma_L, L_0 | x) \propto \mathfrak{L}(x | \gamma_L, L_0) \Pi(\gamma_L, L_0)$, the contour plot for which is shown on Fig.~\ref{fig:posteriorPDF}, with contours in blue and red corresponding respectively to the posterior $90\%$ and $50\%$ credible regions. The constant rate curves shape the posterior: if we fix a value of the rate, higher values for the low-luminosity cutoff $L_0$ favor higher values of the low-luminosity power index $\gamma_L$.  
Each credible region's $L_0$ value is compatible with the luminosity value range of GRB~170817A. Finally, the $90\%$ credible region curve does not close for low values of $L_0$: this is due to the fact that we do not have any information about events down to those luminosities and for this reason we did not explore lower values for $L_0$.
By marginalizing the posterior PDF over $L_0$, we obtain that \gammaL .
\begin{figure}[htbp]
	\begin{center}
		\includegraphics[width=\linewidth]{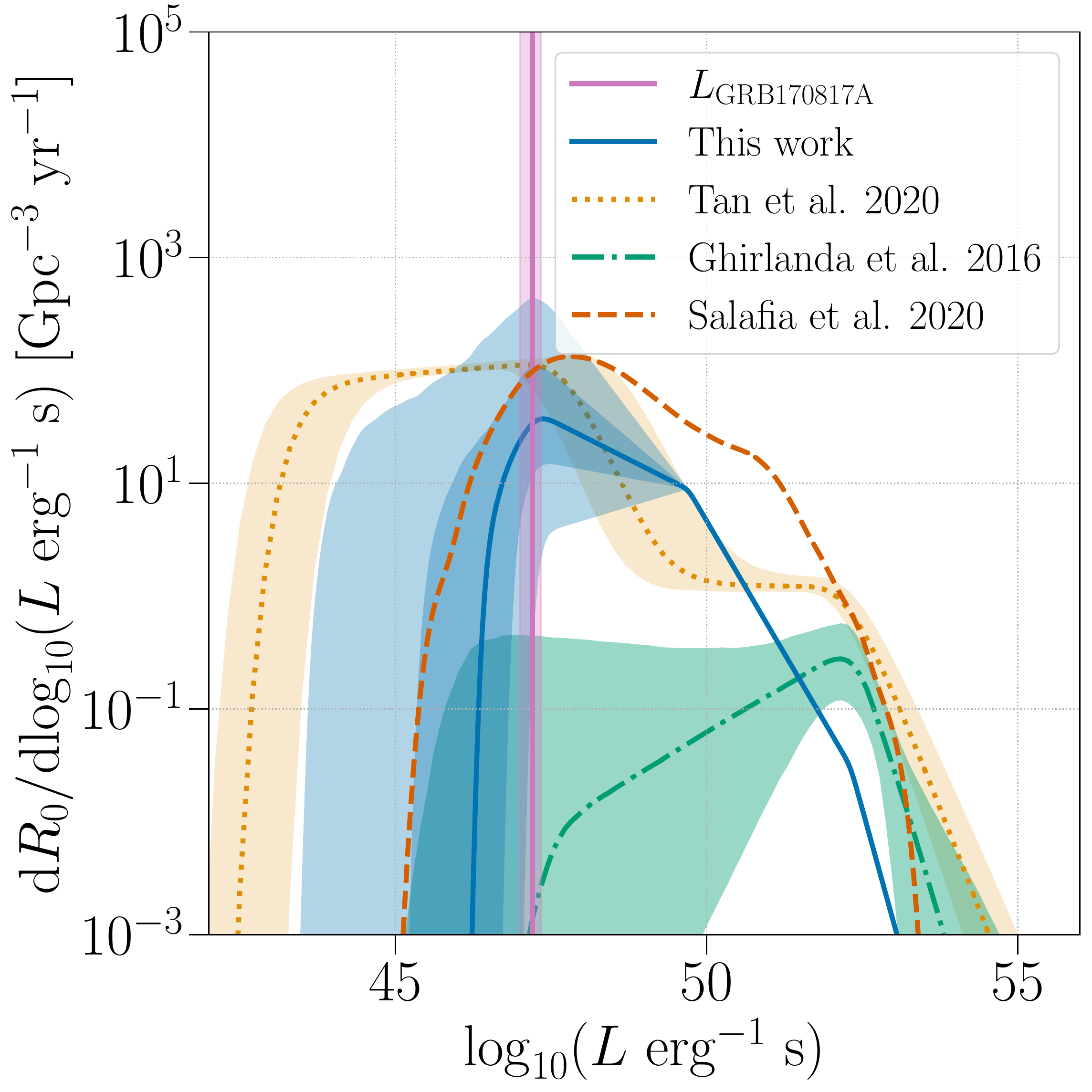}
		\includegraphics[width=\linewidth]{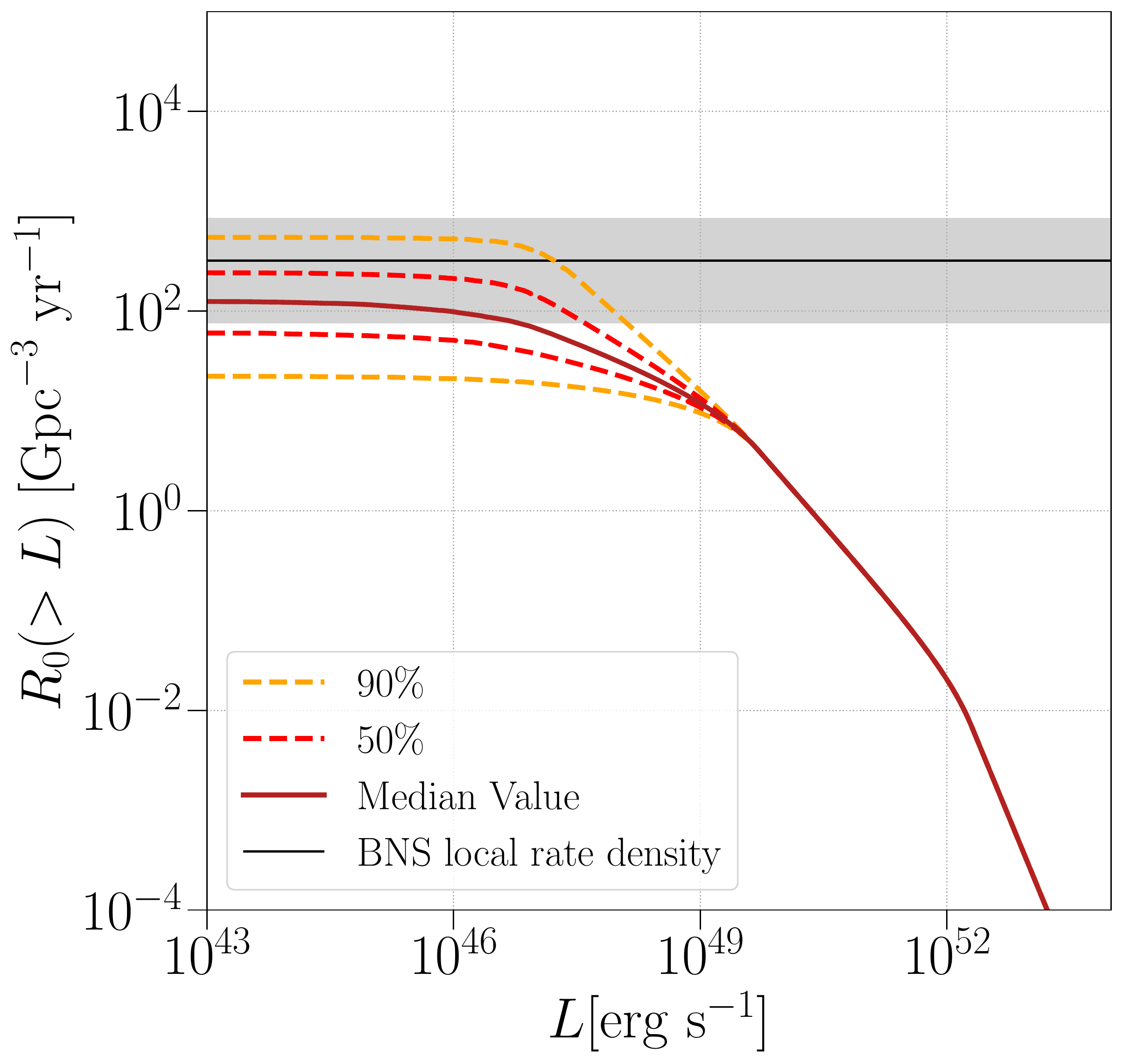}
	\end{center}
	\caption{
		Plots of differential local rate densities as functions of luminosity (upper panel) and of the inverse cumulative rate density as a function of $L$ (lower panel).
		In the upper plot, our set of curves is represented with the blue solid line (the solid line represents the median one and the shaded areas represent the $90\%$ and $50\%$ credible intervals). At large luminosity, we do not represent the error bars, since in this analysis the parameters of the distribution above \Lstarstar were set by the analysis from \citet{Wanderman_2015}. In the same plot we also show the luminosity functions from \protect{\citet{Salafia:2019ldt}} (orange, dashed line), \protect{\citet{Tan:2020vtc}} (yellow, dotted line) and \protect{\citet{Ghirlanda:2016ijf}} (green, dash-dotted line).
		In the bottom plot we represent the median curve as a continuous brown line and the $90\%$ and $50\%$ credible intervals respectively as yellow and red dahsed lines. 	
	}
	\label{fig:dr0_dlogl_R0_maj_L}
\end{figure}

To present these results in the luminosity function space, we compute the rate curves $\mathrm{d} R_0 / \mathrm{d}\log{L}$ for pairs of values $(\gamma_L, L_0)$ sampled according to the posterior distribution $P(\gamma_L, L_0 | x)$. From this set of curves we obtain the median and credible intervals on the luminosity distribution.

The plot in the top panel of Fig.~\ref{fig:dr0_dlogl_R0_maj_L} shows $\mathrm{d} R_0 / \mathrm{d}\log{L}$ $50\%$ credible intervals as functions of $\log{L}$ and compares them to other estimations performed in other works \citep[][]{Salafia:2019ldt, Tan:2020vtc, Ghirlanda:2016ijf}. It illustrates how the short GRB luminosity functions in our model peaks around $L \sim L_{\mathrm{GRB~170817A}}$, considering this the only short GRB event observed at such a low luminosity.

The plot in the bottom panel of Fig.~\ref{fig:dr0_dlogl_R0_maj_L} shows the inverse cumulative short GRB rate density distribution $R_0(>L)$ as a function of the luminosity $L$. The credible intervals corresponding to the sampled curve are compatible with the BNS rate density measured for \citet{Abbott_2021}.

Given the present results on the low-luminosity short GRB population and the expected sensitivity for \aclu{O4} \citep[\ac{O4};][]{Abbott_2020}, and only considering short GRBs detected by Fermi/GBM as onboard triggers, we estimate a joint GW--GRB detection rate of \JDROFour during the next data collecting period.
\section{Conclusions}
\label{sec:conclusions}

We followed-up \emph{Fermi}/{GBM} and \emph{Swift}/{BAT} GRBs reported during LIGO--Virgo's O3b and performed a targeted search using the times of the GRBs and their sky localizations to search for possible GW associations.
For GRBs flagged as either short or ambiguous (see Sec.~\ref{sec:grb_sample}), we ran a template-based search for BNS and NSBH waveforms \citep{Harry:2010fr, Williamson:2014wma}. We also ran on all GRBs a generic transient analysis to look for GW signals \citep{2010NJPh...12e3034S, 2012PhRvD..86b2003W}.
We did not find any significant GW candidate in coincidence with the GRBs we analyzed. Our results are consistent with the previously predicted detection rate of 0.07--1.8 per year for O3 \citep{Authors:2019fue}.
We also performed a weighted binomial test to search for a population of subthreshold GW signals in our sample. We did not find strong evidence for any such event. We used different emission models to put a lower bound on the distances of the GRB progenitors. The 90\% exclusion distances are reported in Table \ref{tab:combined} for all the GRBs in our sample, along with timing and localization information as well as information on detectors used in the analyses. 
Finally, we performed a population study for all GRBs analyzed with the modeled search in O1 \citep{2017ApJ...841...89A}, O2 \citep{Authors:2019fue}, O3a \citep{2020arXiv201014550T} and O3b. Starting from a broken power law to model our population and constraining two of its parameters through Bayesian inference, we found that our luminosity function peaks around the luminosity value measured for GRB~170817A with this model. Furthermore, the local rate density for short GRBs is compatible with that of BNS events. Based on the present population study, we provided an estimate of the joint GW--GRB detection rate for the O4 run.

\acknowledgments
This material is based upon work supported by NSF’s LIGO Laboratory which is a major facility
fully funded by the National Science Foundation.
The authors also gratefully acknowledge the support of
the Science and Technology Facilities Council (STFC) of the
United Kingdom, the Max-Planck-Society (MPS), and the State of
Niedersachsen/Germany for support of the construction of Advanced LIGO 
and construction and operation of the GEO600 detector. 
Additional support for Advanced LIGO was provided by the Australian Research Council.
The authors gratefully acknowledge the Italian Istituto Nazionale di Fisica Nucleare (INFN),  
the French Centre National de la Recherche Scientifique (CNRS) and
the Netherlands Organization for Scientific Research (NWO), 
for the construction and operation of the Virgo detector
and the creation and support  of the EGO consortium. 
The authors also gratefully acknowledge research support from these agencies as well as by 
the Council of Scientific and Industrial Research of India, 
the Department of Science and Technology, India,
the Science \& Engineering Research Board (SERB), India,
the Ministry of Human Resource Development, India,
the Spanish Agencia Estatal de Investigaci\'on (AEI),
the Spanish Ministerio de Ciencia e Innovaci\'on and Ministerio de Universidades,
the Conselleria de Fons Europeus, Universitat i Cultura and the Direcci\'o General de Pol\'{\i}tica Universitaria i Recerca del Govern de les Illes Balears,
the Conselleria d'Innovaci\'o, Universitats, Ci\`encia i Societat Digital de la Generalitat Valenciana and
the CERCA Programme Generalitat de Catalunya, Spain,
the National Science Centre of Poland and the European Union – European Regional Development Fund; Foundation for Polish Science (FNP),
the Swiss National Science Foundation (SNSF),
the Russian Foundation for Basic Research, 
the Russian Science Foundation,
the European Commission,
the European Social Funds (ESF),
the European Regional Development Funds (ERDF),
the Royal Society, 
the Scottish Funding Council, 
the Scottish Universities Physics Alliance, 
the Hungarian Scientific Research Fund (OTKA),
the French Lyon Institute of Origins (LIO),
the Belgian Fonds de la Recherche Scientifique (FRS-FNRS), 
Actions de Recherche Concertées (ARC) and
Fonds Wetenschappelijk Onderzoek – Vlaanderen (FWO), Belgium,
the Paris \^{I}le-de-France Region, 
the National Research, Development and Innovation Office Hungary (NKFIH), 
the National Research Foundation of Korea,
the Natural Science and Engineering Research Council Canada,
Canadian Foundation for Innovation (CFI),
the Brazilian Ministry of Science, Technology, and Innovations,
the International Center for Theoretical Physics South American Institute for Fundamental Research (ICTP-SAIFR), 
the Research Grants Council of Hong Kong,
the National Natural Science Foundation of China (NSFC),
the Leverhulme Trust, 
the Research Corporation, 
the Ministry of Science and Technology (MOST), Taiwan,
the United States Department of Energy,
and
the Kavli Foundation.
The authors gratefully acknowledge the support of the NSF, STFC, INFN and CNRS for provision of computational resources.

This work was supported by MEXT, JSPS Leading-edge Research Infrastructure Program, JSPS Grant-in-Aid for Specially Promoted Research 26000005, JSPS Grant-in-Aid for Scientific Research on Innovative Areas 2905: JP17H06358, JP17H06361 and JP17H06364, JSPS Core-to-Core Program A. Advanced Research Networks, JSPS Grant-in-Aid for Scientific Research (S) 17H06133 and 20H05639 , JSPS Grant-in-Aid for Transformative Research Areas (A) 20A203: JP20H05854, the joint research program of the Institute for Cosmic Ray Research, University of Tokyo, National Research Foundation (NRF) and Computing Infrastructure Project of KISTI-GSDC in Korea, Academia Sinica (AS), AS Grid Center (ASGC) and the Ministry of Science and Technology (MoST) in Taiwan under grants including AS-CDA-105-M06, Advanced Technology Center (ATC) of NAOJ, Mechanical Engineering Center of KEK.

We would like to thank all of the essential workers who put their health at risk during the COVID-19 pandemic, without whom we would not have been able to complete this work.

\begin{longrotatetable}
\begin{deluxetable*}{lcRRlllRRRRR}
  \tablecaption{\label{tab:combined}
    \ac{GRB} details and associated \ac{GW} emission limits for
    each of the {\it Fermi} and {\it Swift} \acp{GRB} followed up on 
    during O3b. The GRB Name column reports each GRB's formal designation \citep{2009GCN.10251....1B}  
    or the Fermi GBM trigger ID when a formal designation has not been assigned. 
    The UTC times reported are rounded to the earlier integer second. 
    The Satellite column gives the satellite that provided the
    \ac{GRB} sky localization used in the \ac{GW} analysis. The  
    Network column lists the \ac{GW} detector network used: H1 = LIGO 
    Hanford, L1 = LIGO Livingston, V1 = Virgo. 
    The $^\dagger$ symbol indicates that the \ac{GRB}'s $T_{90} > 60\,\mathrm{s}$,
    so the generic transient search's on-source window was extended. 
    Where the generic transient search (Sec.\,\ref{sec:Xpipeline-search}) 
    and the modeled search (Sec.\,\ref{sec:cbc-search}) used a different 
    IFO network, the network used by the modeled search is shown in parentheses. 
    The last 5 columns show the 90\% confidence exclusion distances 
    for each \ac{GRB} ($D_{90}$) for the following emission scenarios:
    \ac{BNS}, generic and aligned-spin \ac{NSBH} from the modeled search,
    and from the generic transient search, \ac{ADI}-A, and \ac{CSG}
    \ac{GW} burst at 150\,HZ with total radiated energy $E_{\text{GW}} 
    = 10^{-2}\,\mathrm{\Msun c^2}$.}
  
  \tablewidth{700pt}
  \tabletypesize{\scriptsize}
  \tablehead{
    \colhead{} & \colhead{} & \colhead{} & \colhead{} & \colhead{} &
    \colhead{} & \colhead{} & \multicolumn{5}{c}{$D_{90}$ (Mpc)} \\
    \cmidrule{8-12}
    \colhead{GRB Name} & \colhead{UTC Time} & 
    \colhead{R.A.} & \colhead{Dec.} & 
    \colhead{Satellite} & \colhead{Type} & 
    \colhead{Network} & \colhead{\ac{BNS}} & 
    \colhead{Generic \ac{NSBH}} & \colhead{Aligned \ac{NSBH}} &
    \colhead{ADI-A} & \colhead{CSG 150 Hz}
  } 
  \startdata
  191101A & 21:08:03 & $ 16^{\mathrm{h}} 47^{\mathrm{m}} 25^{\mathrm{s}}$ & $ 43^{\circ} 45' $ & Swift & Long & H1L1V1$^\dagger$  & - & - & - & 204 & 72\\
191106A & 14:15:23 & \hms{17}{57}{26} & \dms{46}{03} & Swift & Ambiguous & H1 & 71 & 112 & 148 & - & -\\
191110A & 14:05:34 & $ 17^{\mathrm{h}} 20^{\mathrm{m}} 48^{\mathrm{s}}$ & $ 43^{\circ} 31' $ & Fermi & Long & H1L1V1  & - & - & - & 104 & 32\\
191111347 & 08:19:09 & $  8^{\mathrm{h}} 42^{\mathrm{m}} 09^{\mathrm{s}}$ & $ -32^{\circ} 28' $ & Fermi & Long & H1L1  & - & - & - & 91 & 31\\
191111A & 08:44:29 & $ 12^{\mathrm{h}} 37^{\mathrm{m}} 09^{\mathrm{s}}$ & $ -32^{\circ} 07' $ & Fermi & Long & H1L1$^\dagger$  & - & - & - & 102 & 42\\
191111B & 13:07:10 & $ 15^{\mathrm{h}} 57^{\mathrm{m}} 38^{\mathrm{s}}$ & $ -70^{\circ} 25' $ & Fermi & Long & H1V1$^\dagger$  & - & - & - & 73 & 26\\
191117A & 00:08:28 & $ 19^{\mathrm{h}} 51^{\mathrm{m}} 31^{\mathrm{s}}$ & $ 76^{\circ} 23' $ & Fermi & Long & H1L1  & - & - & - & 144 & 54\\
191117B & 15:17:38 & $ 10^{\mathrm{h}} 29^{\mathrm{m}} 40^{\mathrm{s}}$ & $ 7^{\circ} 14' $ & Fermi & Ambiguous & H1V1  & 141 & 189 & 257 & 77 & 25\\
191118A & 22:12:01 & $ 14^{\mathrm{h}} 15^{\mathrm{m}} 57^{\mathrm{s}}$ & $ -48^{\circ} 24' $ & Fermi & Long & L1V1  & - & - & - & 78 & 22\\
191119261 & 06:16:07 & $ 20^{\mathrm{h}} 37^{\mathrm{m}} 24^{\mathrm{s}}$ & $ -9^{\circ} 21' $ & Fermi & Long & H1L1  & - & - & - & 39 & 17\\
191122A & 13:32:56 & $  3^{\mathrm{h}} 37^{\mathrm{m}} 09^{\mathrm{s}}$ & $ -32^{\circ} 11' $ & Swift & Long & H1L1V1$^\dagger$  & - & - & - & 148 & 49\\
191123A & 10:38:44 & $ 14^{\mathrm{h}} 21^{\mathrm{m}} 10^{\mathrm{s}}$ & $ 22^{\circ} 50' $ & Swift & Long & L1V1$^\dagger$  & - & - & - & 105 & 32\\
191125A & 04:56:43 & $ 16^{\mathrm{h}} 12^{\mathrm{m}} 21^{\mathrm{s}}$ & $ -13^{\circ} 07' $ & Fermi & Long & H1L1V1$^\dagger$  & - & - & - & 177 & 59\\
191125B & 15:12:45 & $ 23^{\mathrm{h}} 34^{\mathrm{m}} 09^{\mathrm{s}}$ & $ 18^{\circ} 12' $ & Fermi & Long & H1L1V1  & - & - & - & 104 & 35\\
191129A & 03:22:27 & $  0^{\mathrm{h}} 35^{\mathrm{m}} 43^{\mathrm{s}}$ & $ 5^{\circ} 26' $ & Fermi & Long & L1V1  & - & - & - & 69 & 34\\
191130253 & 06:04:41 & $ 23^{\mathrm{h}} 17^{\mathrm{m}} 36^{\mathrm{s}}$ & $ 63^{\circ} 05' $ & Fermi & Long & H1V1$^\dagger$  & - & - & - & 60 & 24\\
191130507 & 12:09:34 & $ 23^{\mathrm{h}} 14^{\mathrm{m}} 24^{\mathrm{s}}$ & $ -7^{\circ} 44' $ & Fermi & Long & L1V1$^\dagger$  & - & - & - & 74 & 36\\
191130A & 13:05:02 & $  8^{\mathrm{h}} 52^{\mathrm{m}} 19^{\mathrm{s}}$ & $ 4^{\circ} 60' $ & Swift & Long & L1V1  & - & - & - & 95 & 33\\
191202A & 20:48:51 & $ 16^{\mathrm{h}} 38^{\mathrm{m}} 08^{\mathrm{s}}$ & $ 17^{\circ} 33' $ & Fermi & Long & H1L1V1  & - & - & - & 183 & 70\\
191203A & 06:57:19 & $ 22^{\mathrm{h}} 09^{\mathrm{m}} 33^{\mathrm{s}}$ & $ 51^{\circ} 49' $ & Fermi & Short & H1L1  & 87 & 153 & 180 & 73 & 32\\
191205741 & 17:46:20 & $  0^{\mathrm{h}} 56^{\mathrm{m}} 09^{\mathrm{s}}$ & $ -34^{\circ} 09' $ & Fermi & Ambiguous & H1L1  & 189 & 211 & 355 & 146 & 58\\
191213254 & 06:05:33 & $ 13^{\mathrm{h}} 04^{\mathrm{m}} 14^{\mathrm{s}}$ & $ -30^{\circ} 27' $ & Fermi & Long & H1L1V1  & - & - & - & 84 & 50\\
191213B & 18:49:07 & $ 22^{\mathrm{h}} 04^{\mathrm{m}} 14^{\mathrm{s}}$ & $ -13^{\circ} 56' $ & Fermi & Long & H1L1V1  & - & - & - & 20 & 9\\
191213A & 04:06:23 & $ 14^{\mathrm{h}} 58^{\mathrm{m}} 07^{\mathrm{s}}$ & $ -9^{\circ} 45' $ & Swift & Long & H1L1V1$^\dagger$  & - & - & - & 39 & 51\\
191220589 & 14:08:29 & $ 14^{\mathrm{h}} 07^{\mathrm{m}} 04^{\mathrm{s}}$ & $ -67^{\circ} 31' $ & Fermi & Long & L1V1  & - & - & - & 74 & 22\\
191220A & 13:29:37 & $ 18^{\mathrm{h}} 45^{\mathrm{m}} 20^{\mathrm{s}}$ & $ 26^{\circ} 40' $ & Swift & Long & L1V1$^\dagger$  & - & - & - & 94 & 33\\
191221A & 19:14:28 & \hms{2}{43}{19} & \dms{-43}{02} & Fermi & Ambiguous & H1V1 & 166 & 219 & 327 & - & -\\
191221B & 20:39:13 & $ 10^{\mathrm{h}} 19^{\mathrm{m}} 19^{\mathrm{s}}$ & $ -38^{\circ} 09' $ & Swift & Long & H1V1$^\dagger$  & - & - & - & 105 & 34\\
191225A & 07:25:16 & $  6^{\mathrm{h}} 21^{\mathrm{m}} 57^{\mathrm{s}}$ & $ -17^{\circ} 21' $ & Fermi & Long & L1V1$^\dagger$  & - & - & - & 92 & 32\\
191225B & 17:37:51 & $  9^{\mathrm{h}} 43^{\mathrm{m}} 12^{\mathrm{s}}$ & $ -7^{\circ} 11' $ & Fermi & Long & H1L1$^\dagger$  & - & - & - & 44 & 32\\
191227723 & 17:21:44 & $ 17^{\mathrm{h}} 12^{\mathrm{m}} 40^{\mathrm{s}}$ & $ -26^{\circ} 01' $ & Fermi & Short & H1L1V1  & 144 & 207 & 253 & 129 & 46\\
191227A & 01:39:37 & $ 21^{\mathrm{h}} 16^{\mathrm{m}} 40^{\mathrm{s}}$ & $ -16^{\circ} 43' $ & Swift & Long & H1V1$^\dagger$  & - & - & - & 98 & 32\\
191228A & 00:01:19 & $  0^{\mathrm{h}} 21^{\mathrm{m}} 27^{\mathrm{s}}$ & $ -8^{\circ} 41' $ & Swift & Long & H1L1V1$^\dagger$  & - & - & - & 148 & 52\\
200101861 & 20:39:26 & $ 17^{\mathrm{h}} 09^{\mathrm{m}} 43^{\mathrm{s}}$ & $ -35^{\circ} 04' $ & Fermi & Long & L1V1  & - & - & - & 70 & 18\\
200103678 & 16:16:50 & $ 23^{\mathrm{h}} 41^{\mathrm{m}} 31^{\mathrm{s}}$ & $ -38^{\circ} 22' $ & Fermi & Long & H1V1  & - & - & - & 36 & 16\\
200103689 & 16:32:23 & $  7^{\mathrm{h}} 53^{\mathrm{m}} 55^{\mathrm{s}}$ & $ -0^{\circ} 54' $ & Fermi & Long & H1L1V1$^\dagger$  & - & - & - & 27 & 18\\
200105914 & 21:55:28 & $ 21^{\mathrm{h}} 32^{\mathrm{m}} 07^{\mathrm{s}}$ & $ -41^{\circ} 11' $ & Fermi & Long & H1L1V1  & - & - & - & 74 & 23\\
200109A & 01:46:16 & $ 20^{\mathrm{h}} 28^{\mathrm{m}} 27^{\mathrm{s}}$ & $ 52^{\circ} 59' $ & Swift & Long & L1V1$^\dagger$  & - & - & - & 80 & 25\\
200110518 & 12:26:08 & $  6^{\mathrm{h}} 24^{\mathrm{m}} 36^{\mathrm{s}}$ & $ 28^{\circ} 53' $ & Fermi & Long & H1V1$^\dagger$  & - & - & - & 60 & 22\\
200112395 & 09:28:27 & $ 12^{\mathrm{h}} 27^{\mathrm{m}} 52^{\mathrm{s}}$ & $ -34^{\circ} 19' $ & Fermi & Long & H1L1V1  & - & - & - & 96 & 36\\
200112A & 12:36:31 & $ 10^{\mathrm{h}} 00^{\mathrm{m}} 31^{\mathrm{s}}$ & $ 64^{\circ} 25' $ & Fermi & Long & H1L1V1  & - & - & - & 141 & 45\\
200114A & 03:40:43 & $ 13^{\mathrm{h}} 17^{\mathrm{m}} 31^{\mathrm{s}}$ & $ -0^{\circ} 19' $ & Fermi & Long & H1L1V1  & - & - & - & 141 & 48\\
200115A & 11:50:23 & $  3^{\mathrm{h}} 45^{\mathrm{m}} 48^{\mathrm{s}}$ & $ 5^{\circ} 36' $ & Swift & Long & H1L1$^\dagger$  & - & - & - & 102 & 34\\
200117517 & 12:24:06 & $  8^{\mathrm{h}} 38^{\mathrm{m}} 40^{\mathrm{s}}$ & $ -62^{\circ} 31' $ & Fermi & Long & H1L1V1  & - & - & - & 99 & 27\\
200120A & 23:04:55 & $  9^{\mathrm{h}} 08^{\mathrm{m}} 32^{\mathrm{s}}$ & $ -70^{\circ} 26' $ & Fermi & Long & H1V1  & - & - & - & 105 & 46\\
200122221 & 05:18:20 & $  8^{\mathrm{h}} 18^{\mathrm{m}} 38^{\mathrm{s}}$ & $ 67^{\circ} 05' $ & Fermi & Ambiguous & H1L1V1  & 183 & 247 & 371 & 162 & 41\\
200122A & 01:41:00 & $ 14^{\mathrm{h}} 00^{\mathrm{m}} 02^{\mathrm{s}}$ & $ 27^{\circ} 33' $ & Swift & Long & H1L1V1$^\dagger$  & - & - & - & 101 & 32\\
200125B & 20:43:31 & $  0^{\mathrm{h}} 29^{\mathrm{m}} 47^{\mathrm{s}}$ & $ 64^{\circ} 41' $ & Fermi & Long & H1L1  & - & - & - & 176 & 69\\
200126466 & 11:10:51 & $  3^{\mathrm{h}} 57^{\mathrm{m}} 52^{\mathrm{s}}$ & $ -59^{\circ} 37' $ & Fermi & Short & L1V1  & 149 & 214 & 300 & 102 & 23\\
200127B & 18:11:18 & $  5^{\mathrm{h}} 03^{\mathrm{m}} 33^{\mathrm{s}}$ & $ 20^{\circ} 04' $ & Fermi & Long & H1L1  & - & - & - & 95 & 33\\
200128A & 03:40:05 & $ 10^{\mathrm{h}} 34^{\mathrm{m}} 36^{\mathrm{s}}$ & $ 41^{\circ} 34' $ & Fermi & Short & L1V1  & 149 & 207 & 253 & 94 & 27\\
200129A & 09:48:44 & $ 23^{\mathrm{h}} 05^{\mathrm{m}} 07^{\mathrm{s}}$ & $ -44^{\circ} 58' $ & Fermi & Short & H1L1V1(H1L1)  & 235 & 323 & 454 & 203 & 64\\
200130A & 05:57:16 & $ 21^{\mathrm{h}} 57^{\mathrm{m}} 09^{\mathrm{s}}$ & $ -65^{\circ} 56' $ & Fermi & Long & H1V1  & - & - & - & 110 & 34\\
200130B & 09:59:56 & $  9^{\mathrm{h}} 10^{\mathrm{m}} 07^{\mathrm{s}}$ & $ -51^{\circ} 20' $ & Fermi & Long & H1L1  & - & - & - & 14 & 6\\
200131A & 22:41:15 & $  0^{\mathrm{h}} 12^{\mathrm{m}} 21^{\mathrm{s}}$ & $ 51^{\circ} 07' $ & Swift & Long & H1L1V1  & - & - & - & 213 & 78\\
200201A & 00:57:20 & $ 19^{\mathrm{h}} 10^{\mathrm{m}} 50^{\mathrm{s}}$ & $ -11^{\circ} 02' $ & Fermi & Long & H1L1V1  & - & - & - & 102 & 34\\
200205C & 20:17:23 & $ 14^{\mathrm{h}} 52^{\mathrm{m}} 50^{\mathrm{s}}$ & $ -42^{\circ} 47' $ & Fermi & Long & H1L1  & - & - & - & 142 & 50\\
200207A & 01:22:55 & $ 16^{\mathrm{h}} 16^{\mathrm{m}} 21^{\mathrm{s}}$ & $ -48^{\circ} 18' $ & Fermi & Long & L1V1  & - & - & - & 71 & 32\\
200208A & 01:14:17 & $  1^{\mathrm{h}} 46^{\mathrm{m}} 55^{\mathrm{s}}$ & $ 25^{\circ} 19' $ & Fermi & Long & L1V1  & - & - & - & 95 & 46\\
200211A & 07:26:28 & $ 23^{\mathrm{h}} 00^{\mathrm{m}} 48^{\mathrm{s}}$ & $ -7^{\circ} 09' $ & Fermi & Long & H1L1$^\dagger$  & - & - & - & 188 & 72\\
200212A & 10:49:49 & $  8^{\mathrm{h}} 19^{\mathrm{m}} 11^{\mathrm{s}}$ & $ 22^{\circ} 57' $ & Fermi & Long & H1L1  & - & - & - & 94 & 34\\
200215A & 14:39:31 & $  2^{\mathrm{h}} 16^{\mathrm{m}} 24^{\mathrm{s}}$ & $ 12^{\circ} 47' $ & Swift & Long & H1V1  & - & - & - & 101 & 33\\
200216A & 09:07:25 & $ 20^{\mathrm{h}} 45^{\mathrm{m}} 45^{\mathrm{s}}$ & $ -11^{\circ} 39' $ & Swift & Long & H1L1V1  & - & - & - & 155 & 51\\
200216B & 13:32:33 & $ 10^{\mathrm{h}} 41^{\mathrm{m}} 49^{\mathrm{s}}$ & $ 19^{\circ} 28' $ & Swift & Long & H1L1V1$^\dagger$  & - & - & - & 93 & 33\\
200219B & 09:54:14 & $ 19^{\mathrm{h}} 56^{\mathrm{m}} 31^{\mathrm{s}}$ & $ 6^{\circ} 39' $ & Fermi & Long & H1L1V1  & - & - & - & 122 & 46\\
200219A & 07:36:49 & $ 22^{\mathrm{h}} 50^{\mathrm{m}} 22^{\mathrm{s}}$ & $ -59^{\circ} 06' $ & Swift & Long & H1L1V1$^\dagger$  & - & - & - & 215 & 76\\
200221A & 03:52:58 & $ 10^{\mathrm{h}} 28^{\mathrm{m}} 24^{\mathrm{s}}$ & $ 33^{\circ} 08' $ & Fermi & Ambiguous & H1L1V1  & 213 & 247 & 377 & 152 & 55\\
200223A & 19:32:03 & $ 16^{\mathrm{h}} 33^{\mathrm{m}} 33^{\mathrm{s}}$ & $ -55^{\circ} 47' $ & Fermi & Long & L1V1$^\dagger$  & - & - & - & 102 & 33\\
200224B & 05:05:49 & $ 11^{\mathrm{h}} 51^{\mathrm{m}} 36^{\mathrm{s}}$ & $ -28^{\circ} 52' $ & Fermi & Long & H1L1V1$^\dagger$  & - & - & - & 94 & 34\\
200224C & 09:58:44 & $ 12^{\mathrm{h}} 28^{\mathrm{m}} 04^{\mathrm{s}}$ & $ -19^{\circ} 33' $ & Fermi & Short & H1L1V1  & 120 & 160 & 220 & 112 & 33\\
200224A & 03:24:49 & $ 16^{\mathrm{h}} 34^{\mathrm{m}} 58^{\mathrm{s}}$ & $ 41^{\circ} 40' $ & Swift & Long & H1L1  & - & - & - & 10 & 7\\
200227A & 07:20:08 & $  3^{\mathrm{h}} 45^{\mathrm{m}} 43^{\mathrm{s}}$ & $ 9^{\circ} 29' $ & Swift & Long & H1L1V1  & - & - & - & 69 & 21\\
200228A & 06:58:33 & $ 21^{\mathrm{h}} 52^{\mathrm{m}} 02^{\mathrm{s}}$ & $ -46^{\circ} 27' $ & Fermi & Ambiguous & H1L1V1(H1L1)  & 282 & 399 & 528 & 231 & 76\\
200228B & 11:14:41 & $ 16^{\mathrm{h}} 48^{\mathrm{m}} 01^{\mathrm{s}}$ & $ 16^{\circ} 58' $ & Swift & Long & H1L1  & - & - & - & 203 & 73\\
200301320 & 07:40:46 & $ 21^{\mathrm{h}} 20^{\mathrm{m}} 33^{\mathrm{s}}$ & $ 7^{\circ} 30' $ & Fermi & Long & H1L1V1  & - & - & - & 155 & 48\\
200303A & 02:34:57 & $ 14^{\mathrm{h}} 10^{\mathrm{m}} 47^{\mathrm{s}}$ & $ 51^{\circ} 22' $ & Swift & Long & H1L1V1$^\dagger$  & - & - & - & 100 & 30\\
200306B & 22:25:25 & $ 23^{\mathrm{h}} 18^{\mathrm{m}} 07^{\mathrm{s}}$ & $ 5^{\circ} 06' $ & Fermi & Ambiguous & L1V1  & 98 & 156 & 221 & 28 & 19\\
200306C & 22:50:39 & $ 13^{\mathrm{h}} 14^{\mathrm{m}} 18^{\mathrm{s}}$ & $ 11^{\circ} 16' $ & Swift & Long & L1V1$^\dagger$  & - & - & - & 101 & 32\\
200307A & 21:26:59 & $  5^{\mathrm{h}} 36^{\mathrm{m}} 43^{\mathrm{s}}$ & $ -46^{\circ} 13' $ & Fermi & Ambiguous & H1L1(H1L1V1)  & 121 & 172 & 195 & 96 & 34\\
200308A & 22:35:15 & $ 21^{\mathrm{h}} 49^{\mathrm{m}} 45^{\mathrm{s}}$ & $ -27^{\circ} 45' $ & Fermi & Long & H1L1V1  & - & - & - & 124 & 39\\
200311A & 15:16:12 & $ 13^{\mathrm{h}} 35^{\mathrm{m}} 57^{\mathrm{s}}$ & $ -49^{\circ} 41' $ & Fermi & Long & H1L1  & - & - & - & 133 & 49\\
200313A & 01:41:36 & $  5^{\mathrm{h}} 03^{\mathrm{m}} 09^{\mathrm{s}}$ & $ 20^{\circ} 53' $ & Fermi & Long & H1L1V1  & - & - & - & 210 & 73\\
200313B & 10:57:12 & \hms{13}{28}{41} & \dms{40}{30} & Fermi & Ambiguous & L1V1 & 181 & 219 & 286 & - & -\\
200317A & 00:40:30 & $  4^{\mathrm{h}} 22^{\mathrm{m}} 09^{\mathrm{s}}$ & $ -46^{\circ} 19' $ & Fermi & Long & H1L1V1  & - & - & - & 54 & 16\\
200319A & 07:44:40 & $  4^{\mathrm{h}} 21^{\mathrm{m}} 09^{\mathrm{s}}$ & $ -21^{\circ} 15' $ & Fermi & Long & H1L1  & - & - & - & 137 & 47\\
200320A & 09:56:46 & $ 12^{\mathrm{h}} 17^{\mathrm{m}} 21^{\mathrm{s}}$ & $ -38^{\circ} 44' $ & Fermi & Long & H1L1V1  & - & - & - & 44 & 15\\
200323006 & 00:08:42 & $ 22^{\mathrm{h}} 56^{\mathrm{m}} 50^{\mathrm{s}}$ & $ 53^{\circ} 02' $ & Fermi & Long & L1V1  & - & - & - & 62 & 17\\
200323A & 18:46:32 & \hms{10}{25}{53} & \dms{-55}{32} & Fermi & Ambiguous & V1 & 19 & 25 & 38 & - & -\\
200326A & 12:24:47 & $ 16^{\mathrm{h}} 21^{\mathrm{m}} 19^{\mathrm{s}}$ & $ -21^{\circ} 05' $ & Fermi & Long & H1L1V1$^\dagger$  & - & - & - & 104 & 33\\
  \enddata
\end{deluxetable*}
\end{longrotatetable}

\bibliographystyle{aasjournal}

\end{document}